\documentclass[a4paper,11pt]{article}

\usepackage{amsmath,amssymb,epsfig}

\begin{document}	

\title{Predictability, Risk and Online Management in a Complex System of Adaptive Agents}
\author{David M.D. Smith and Neil F. Johnson\\ Physics Department, Oxford University, Oxford OX1 3PU, U.K.}

\date{\today}
\maketitle
\begin{abstract}
We discuss the feasibility of predicting, managing and subsequently manipulating, the future evolution of a Complex Adaptive System. Our archetypal system mimics a population of adaptive, interacting objects, such as those arising in the domains of human health and biology (e.g. cells),  financial markets (e.g. traders), and mechanical systems (e.g. robots).   We show that 
short-term prediction yields corridors along which the model system will, with high probability, evolve. We show how the widths and average direction of these corridors varies in time as the system passes through regions, or {\em pockets}, of enhanced predictability and/or risk. We then show how small 
amounts of `population engineering' can be undertaken in order to steer the system away from any undesired regimes which have been 
predicted. Despite the system's many degrees of freedom and inherent stochasticity, this dynamical, `soft' control over future 
risk requires only minimal knowledge about the underlying composition of the constituent multi-agent population.
\end{abstract}

\section{Introduction}\label{sec:intro}
Complex Adaptive Systems (CAS) are of great theoretical interest because 
they comprise large numbers of interacting objects or `agents' which, unlike particles in traditional physics, change their 
behaviour based on experience \cite{bocc}. Such adaptation yields complicated feedback processes at the microscopic level, which 
in turn generate complicated global dynamics at the macroscopic level.   CAS also arguably represent the `hard' problem in  
biology, engineering, computation and sociology\cite{bocc}. Depending on the application domain, the agents in CAS may be taken 
as representing species, people, cells, computer hardware or software, and are typically quite numerous, e.g. 
$10^2-10^3$\cite{bocc,Wolp}. 

There is also great practical interest in the problem of predicting and subsequently controlling a Complex Adaptive System. 
Consider the enormous task facing a Complex Adaptive Systems `manager' in charge of overseeing some complicated computational, 
biological, medical, sociological or even  economic system. He would certainly like to be able to predict its future evolution 
with sufficient accuracy that he could foresee the system heading towards any `dangerous' areas. However, prediction is not 
enough -- he also needs to be able to steer the system away from this dangerous regime. Furthermore, the CAS manager needs to be 
able to achieve this $without$ detailed knowledge of the present state of its thousand different components, nor does he want to 
have to shut down the system completely. Instead he is seeking for some form of `soft' control which can be applied `online' while the system is still evolving. 

Such online management of a Complex System  therefore represents a significant theoretical and practical challenge. However, the motivation for pursuing such a goal is equally great given the wide range of important real-world systems that can be regarded as complex -- from engineering systems through to human health, social systems and even financial systems. 

In purely deterministic 
systems with only a few degrees of freedom, it is well known that highly complex dynamics such as chaos can arise 
\cite{strogatz} making any control very difficult. The `butterfly effect' whereby small perturbations can have huge 
uncontrollable consequences, comes to mind. One would think that things would be considerably worse in  a CAS, given the much 
larger number of interacting objects. As an additional complication, a CAS may also contain stochastic processes at the 
microscopic and/or macroscopic levels, thereby adding an inherently random element to the system's dynamical evolution.  The 
Central Limit Theorem tells us that the combined effect of a large number of stochastic processes tends fairly rapidly to a 
Gaussian distribution. Hence, one would think that even with reasonably complete knowledge of the present and past states of the 
system, the evolution would be essentially diffusive and hence difficult to control without imposing substantial global 
constraints.

In this paper, we address this question of dynamical control for a simplified yet highly non-trivial model of a CAS. We show 
that a surprising level of prediction and subsequent control can be achieved by introducing small perturbations to the agent 
heterogeneity, i.e. `population engineering'. In particular, the system's global evolution can be managed and undesired future 
scenarios avoided. Despite the many degrees of freedom and inherent stochasticity both at the microscopic and macroscopic 
levels, this global control requires only minimal knowledge on the part of the `system manager'. For the somewhat simpler case 
of Cellular Automata, Israeli and Goldenfeld\cite{golden} have recently obtained the remarkable result that computationally 
irreducible physical processes can become computationally reducible at a coarse-grained level of description. Based on our 
findings, we speculate that similar ideas may hold for a far wider class of system comprising populations of decision-taking, 
adaptive agents. 

It is widely believed (see for example, Ref. \cite{casti1}) that Arthur's so-called El Farol Bar Problem \cite{arthur,A258} 
provides a representative toy model of a  CAS where objects, components or individuals compete for some limited global resource 
(e.g. space in an overcrowded area). To make this model more complete in terms of real-world complex systems, the effect of 
network interconnections has recently been incorporated \cite{nets1, nets2, nets3, nets4}. The El Farol Bar Problem concerns the 
collective decision-making of a group of potential bar-goers (i.e. agents) who use limited global information to predict whether 
they should attend a potentially overcrowded bar on a given night each week. The Statistical Mechanics community has adopted a 
binary version of this problem, the so-called Minority Game (MG) (see Refs.\cite{A246} through \cite{PRE65}), as a new form of 
Ising model which is worthy of study in its own right because of its highly non-trivial dynamics. Here we consider a general 
version of such multi-agent games which (a) incorporates a finite time-horizon $H$ over which agents remember their strategies' 
past successes, to reflect the fact that the more recent past should have more influence than the distant past, and (b)  allows 
for  fluctuations in agent numbers, since agents only participate if they possess a strategy with a sufficiently high success 
rate\cite{thmg}. The formalism we employ is applicable to any CAS which can be mapped onto a population of $N$ objects 
repeatedly taking actions in the form of some global `game'.

The paper has several parts. Initially (section~\ref{sec:Dice}), we discuss a very simple two state `game' to introduce and 
familiarise the reader to the nature of the mathematics which is explored in further detail in the rest of the paper. We then 
more formally establish a common framework for describing the spectrum of future paths of the complex adaptive system 
(section~\ref{sec:formal}). This framework
is general to any complex system which can be mapped onto a general B-A-R (Binary Agent Resource)
model in which the system's future evolution is governed by past history over an arbitrary but
finite time window $H$ (the $Time-Horizon$). In fact, this formalism can be applied to any CAS whose internal dynamics are 
governed by a Markov Process\cite{thmg},  providing the tools whereby we can monitor the future
evolution both with and without the perturbations to the population's composition. In section~\ref{sec:BAR},
we discuss the B-A-R model in more detail, further information is provided in \cite{BAR}. We emphasize that such B-A-R systems 
are {\em not}
limited to the well-known El Farol Bar Problem and Minority Games  - instead these two examples are specific limiting
cases. Initial investigations of a finite time-horizon version of the
Minority Game were first presented in \cite{thmg}. In section~\ref{sec:natural}, we consider the system's evolution in the 
absence of any such
perturbations, hence representing the system's natural evolution. In section~\ref{sec:evolution}, we  revisit this evolution in 
the
presence of control, where this control is limited to relatively minor perturbations at the level
of the heterogeneity of the population. In section~\ref{sec:reduction} we revisit the toy model of section~\ref{sec:Dice} to 
provide a reduced form of formalism for generating averaged quantities of the future possibilities.
In section~\ref{sec:conclusion} we discuss concluding remarks and possible extensions.

\section{A Tale of Two Dice}\label{sec:Dice}
In this section we examine a very simple toy model employed to generate a time series (analogue output) and introduce the 
Future-Cast formalism to describe the model's properties. This toy model comprises two internal states, \textbf{\emph{A}} and 
\textbf{\emph{B}},  and two dice also denoted \textbf{\emph{A}} and \textbf{\emph{B}}. We make these dice generic in that we 
assign their faces values and these are not equal in likelihood.
The rules of the model are very simple. When the system is in state  \textbf{\emph{A}}, dice  \textbf{\emph{A}} is rolled and 
similarly for dice \textbf{\emph{B}}.
The outcome, $\delta_t$, of the relevant dice is used to increment a time (price) series, whose update can be written
\begin{equation}
\label{eq:PriceInc}
S_{t+1} = S_t +  \delta_t 
\end{equation}

\begin{figure}[h]
\begin{center}
\includegraphics[width=0.65\textwidth]{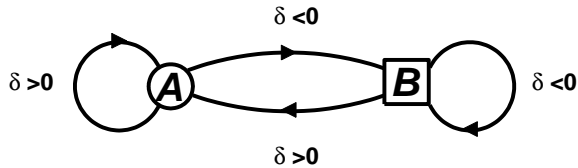}
\caption{ \label{fig:DiceTransitions} The internal state transitions on the De Bruijn graph.}
\end{center} 
\end{figure}
The model employs a very simple rule to govern the transitions between its internal states. If the outcome  $\delta_t$
is greater than zero (recall that we have re-assigned the values on the faces) the internal state at time $t+1$ is 
\textbf{\emph{A}}  and consequently dice \textbf{\emph{A}}  will be used at the next step regardless of the dice used at this 
time step. Conversely, if $\delta_t~<~0$, the internal state at $t+1$ will be \textbf{\emph{B}} and dice \textbf{\emph{B}} used 
for the next increment\footnote{This state transition rule is not critical to the formalism. For example, the rule could be to 
use dice \textbf{\emph{A}} if the last increment was odd and \textbf{\emph{B}} is even, or indeed any property of the outcomes. 
}. These transitions are shown in figure~\ref{fig:DiceTransitions}.

\begin{figure}[h]
\begin{center}
\includegraphics[width=0.75\textwidth]{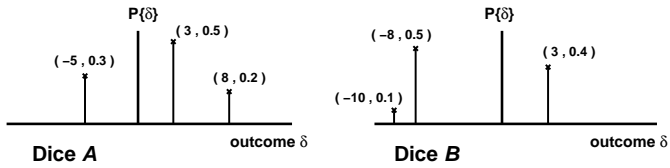}
\caption{ \label{fig:DiceProbs} The outcomes and associated probabilities of our two dice.}
\end{center} 
\end{figure}
\begin{figure}[h]
\begin{center}
\includegraphics[width=0.75\textwidth]{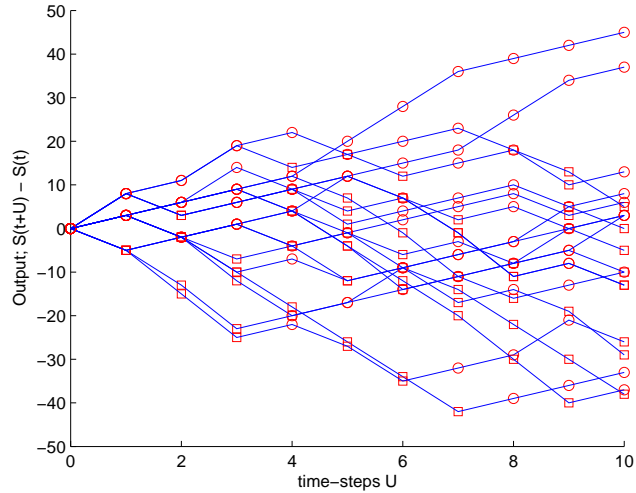}
\caption{ \label{fig:output} The toy model's output over 10 time-steps for 20 realisations where the state at time $t$ is  
\textbf{\emph{A}} for all paths.. The circles represent the system being in state \textbf{\emph{A}}  and the squares the system 
in \textbf{\emph{B}}. }
\end{center} 
\end{figure}

Let us prescribe our dice some values and observe the output of the system, namely rolling dice \textbf{\emph{A}} 
could yield values $\{-5,~+3,~ +8 \}$ with probabilities $\{ 0.3,~ 0.5,~ 0.2\}$ and \textbf{\emph{B}} yields values
$\{-10,~-8,~ +3 \}$ with probabilities $\{ 0.1,~ 0.5,~ 0.4\}$. These are shown in figure~\ref{fig:DiceProbs}.
Let us consider that at some time $t$ the system is in state \textbf{\emph{A}} with the value of the time series being $S(t)$ 
and we wish to investigate the possible output over the next $U$ time-steps ($S(t+U)$). Some examples of the system's change in 
output over the next $10$ time-steps are shown in figure~\ref{fig:output}. The circles represent the system being in state 
\textbf{\emph{A}}  and the squares the system in \textbf{\emph{B}}. 

\begin{figure}[h]
\begin{center}
\includegraphics[width=0.75\textwidth]{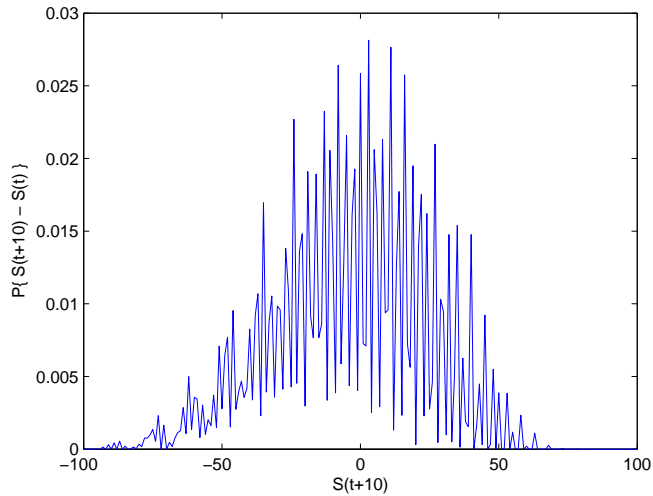}
\caption{ \label{fig:output2} The change in the system's output $S(t+U)~-~S(t)$  at  $U = 10$, and associated probability as 
calculated for $10^6$ time-series realisations.}
\end{center} 
\end{figure}

The stochasticity inherent in the system is evident in figure~\ref{fig:output}. Many possible paths could be realised even 
though they all originate from
state \textbf{\emph{A}}, and only a few of them are depicted. If one wanted to know more about system's output after these $10$ 
time-steps, one could look at many  more runs and look at a histogram of the output for $U~=10$. This Monte-Carlo\cite{Monte} 
technique has been carried out in figure~\ref{fig:output2}. It denotes the possible change in value of the systems analogue 
output after $10$ time-steps and associated probability derived from  $10^6$ runs of the system all with the same initial 
starting state (\textbf{\emph{A}}) at time $t$. However, this method is a numerical approximation. For accurate analysis of the 
system over longer time scales, or for more complex systems, it might prove both inaccurate and/or computationally intensive. 

\begin{figure}[h]
\begin{center}
\includegraphics[width=0.75\textwidth]{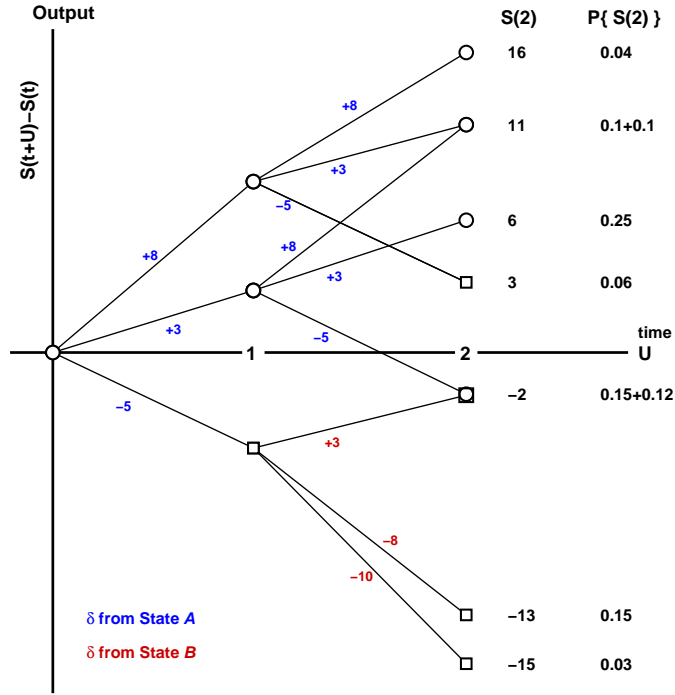}
\caption{ \label{fig:2step} All possible paths after $2$ time-steps and associated probability.}
\end{center} 
\end{figure}

\begin{figure}[h]
\begin{center}
\includegraphics[width=0.95\textwidth]{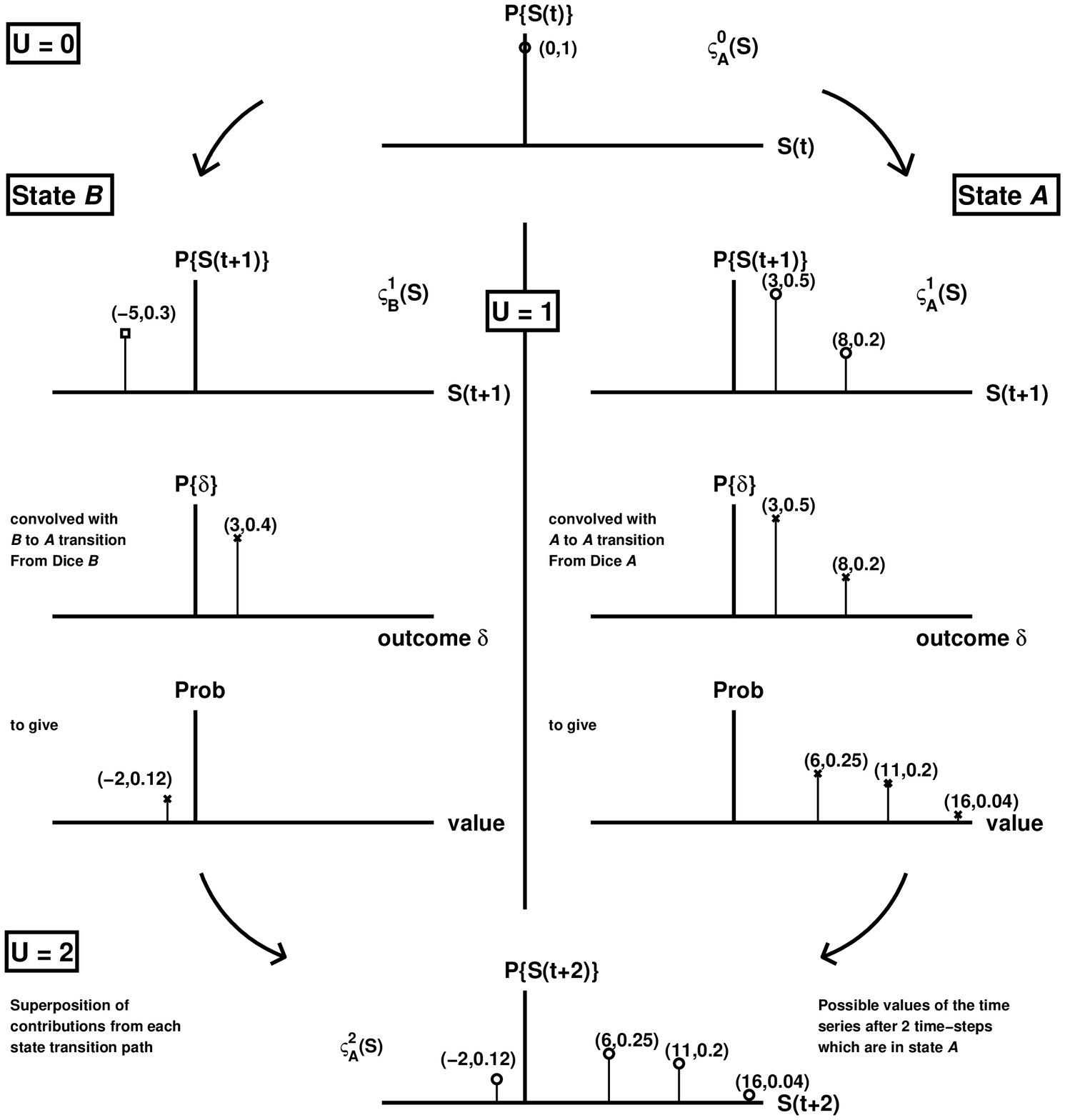}
\caption{ \label{fig:conv} All possible paths after $2$ time-steps which could result in the system being in state 
\textbf{\emph{A}}.}
\end{center} 
\end{figure}

For a more detailed analysis of our model, we must look at the internal state dynamics and how they map to the output. Let us 
consider all possible eventualities over $2$ time-steps, again starting in state \textbf{\emph{A}} at time $t$.
All possible paths are denoted on figure~\ref{fig:2step}. The resulting possible values of change in the output at time $t+2$ 
time-steps and their associated probabilities are given explicitly too. These values are exact in that they are calculated from 
the dice themselves. The $Future-Cast$ frame work which we now introduce will allow us to perform similar analysis over much 
longer periods.
First consider the possible values of $S(t+2)-S(t)$, which result in the system being in state \textbf{\emph{A}}. The state 
transitions that could have 
occurred for these to arise are \textbf{\emph{A}} $\rightarrow$ \textbf{\emph{A}} $\rightarrow$ \textbf{\emph{A}} or 
\textbf{\emph{A}} $\rightarrow$ \textbf{\emph{B}} $\rightarrow$ \textbf{\emph{A}}.  The paths following the former can be 
considered a convolution (explores all possible paths and probabilities\footnote{\label{foot:conv}We define and use the discrete 
convolution operator $\otimes$ such that $(f\otimes g)\mid_i~=~\sum_{j = -\infty}^\infty f(i-j)g(i)$. }) of the distribution of 
possible values of $S(t+1)-S(t)$ in state \textbf{\emph{A}} with  the distribution corresponding to the \textbf{\emph{A}} 
$\rightarrow$ \textbf{\emph{A}} transition. Likewise, the latter a convolution of the distribution of $S(t+1)-S(t)$ in state 
\textbf{\emph{B}} with  the distribution corresponding to a \textbf{\emph{B}} $\rightarrow$ \textbf{\emph{A}} transition. The 
resultant distribution of possibilities in state \textbf{\emph{A}} at time $t+2$ is just the superposition of these convolutions 
as described in figure~\ref{fig:conv}. We can set the initial value $S(t)$ to zero because the absolute position has no bearing 
on the evolution on the system. As such $S(t+U)-S(t)~\equiv S(t+U)$.

\begin{figure}[h]
\begin{center}
\includegraphics[width=0.75\textwidth]{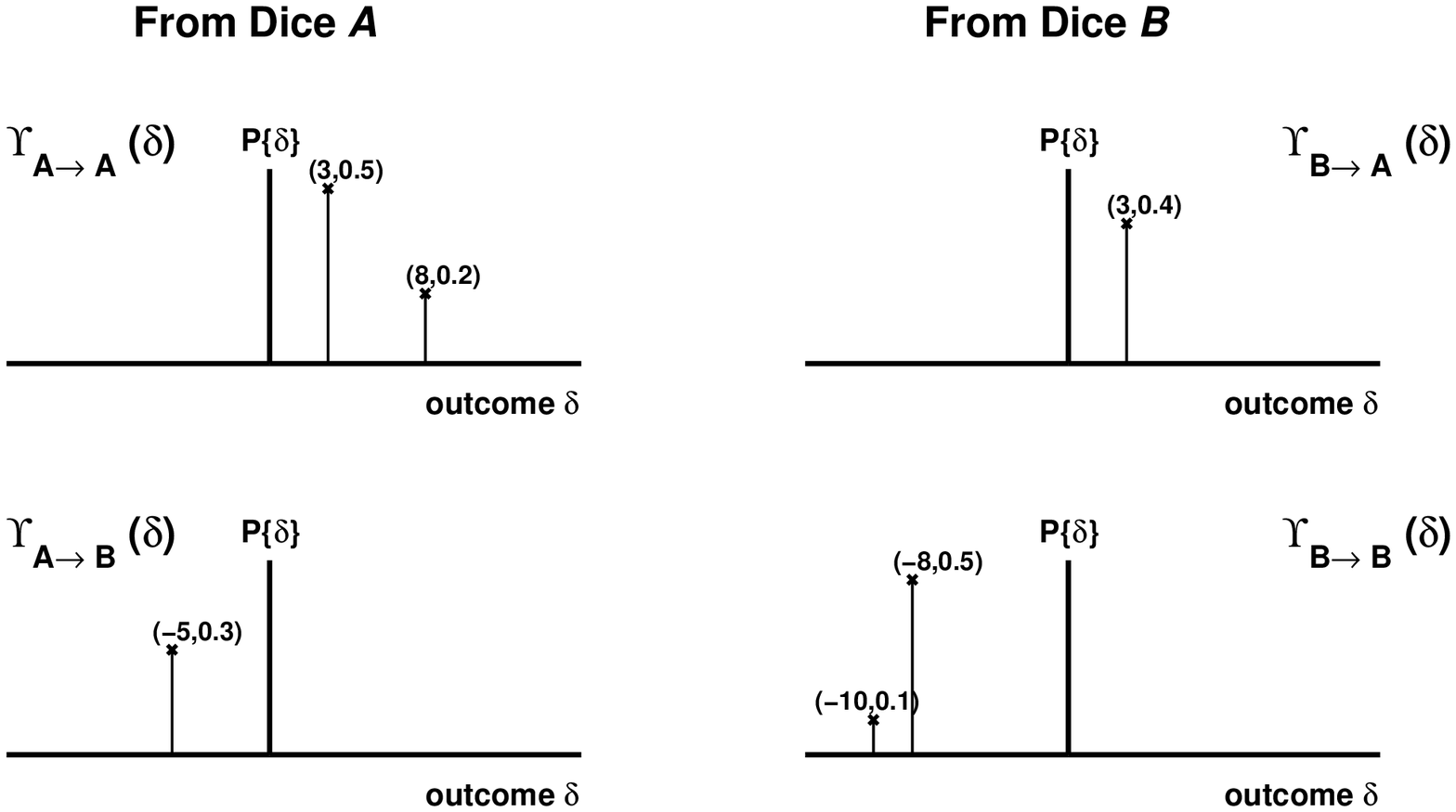}
\caption{ \label{fig:dicetrans} The state transition distributions as prescribed by our dice.}
\end{center} 
\end{figure}

So we note that if at some time $t+U$ there exists a distribution of values in our output which are in state \textbf{\emph{A}} 
then the convolution of this with the \textbf{\emph{B}} $\rightarrow$ \textbf{\emph{A}} transition distribution will result in a 
contribution at $t+U+1$ to our distribution of $S$ which will also be in  \textbf{\emph{A}}. This is evident in the right hand 
side of \ref{fig:conv}. Let us define all these distributions. We denote all possible  values in our output $U$ time-steps 
beyond the present (time $t$) that are  in state  \textbf{\emph{A}}  as the function $\varsigma^U_A(S)$ and $\varsigma^U_B(S)$ 
is the function that describes the values of our output time series  that are in state \textbf{\emph{B}} (as shown in 
figure~\ref{fig:conv}). We can similarly describe the transitions as prescribed by our dice. The possible values allowed and 
corresponding likelihoods for the  \textbf{\emph{A}} $\rightarrow$ \textbf{\emph{A}} transition are denoted $\Upsilon_{A\to 
A}(\delta)$ and similarly $\Upsilon_{A\to B}(\delta)$ for  \textbf{\emph{A}} $\rightarrow$ \textbf{\emph{B}}, $\Upsilon_{B\to 
A}(\delta)$ for  \textbf{\emph{B}} $\rightarrow$ \textbf{\emph{A}} and $\Upsilon_{B\to B}(\delta)$ for  the \textbf{\emph{B}} 
$\rightarrow$ \textbf{\emph{B}} state change. These are shown in figure~\ref{fig:dicetrans}.

\begin{figure}[h]
\begin{center}
\includegraphics[width=0.75\textwidth]{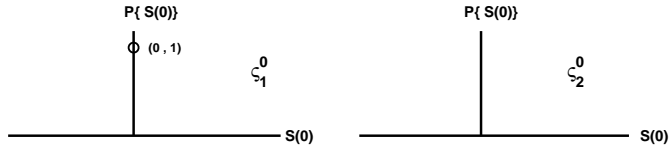}
\caption{ \label{fig:start1} The initial elements of our output distributions vector, $\underline{\varsigma^{0}}$ when starting 
the system in state \textbf{\emph{A}} with initial value of zero in our time series.}
\end{center} 
\end{figure}

\begin{figure}[h]
\begin{center}
\includegraphics[width=0.75\textwidth]{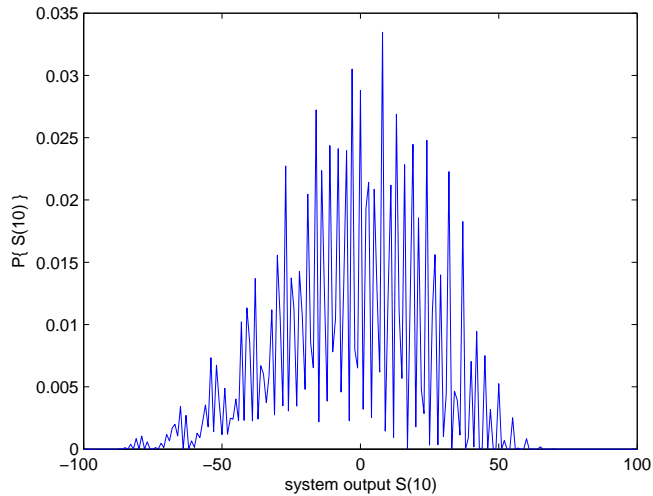}
\caption{ \label{fig:dicefutcast} The actual probability distribution $\Pi^{10} (S))$ of the output after 10 time-steps starting 
the system in state \textbf{\emph{A}} with initial value of zero in our time series.}
\end{center} 
\end{figure}

\begin{figure}[h]
\begin{center}
\includegraphics[width=0.75\textwidth]{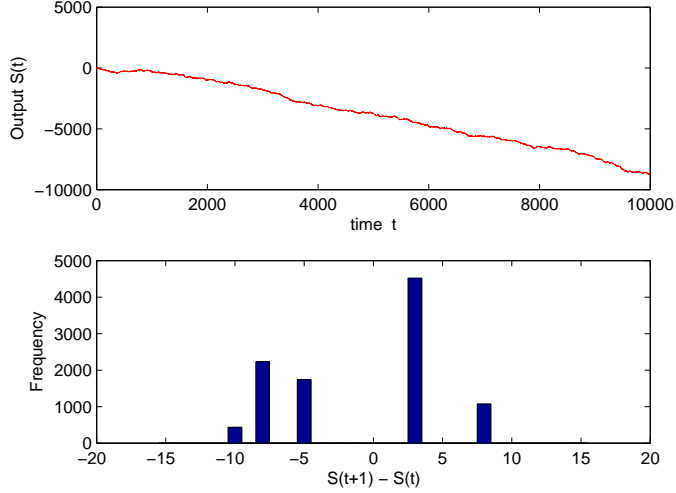}
\caption{ \label{fig:hist1} The one step increments of our time-series as run over 10000 steps.}
\end{center} 
\end{figure}

\begin{figure}[h]
\begin{center}
\includegraphics[width=0.5\textwidth]{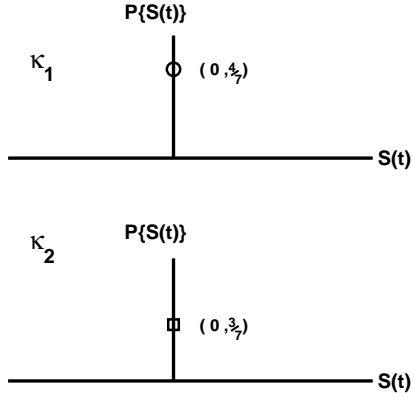}
\caption{ \label{fig:kappa} Explicit depiction of the elements of vector $\underline{\kappa}$ which is used to analyse 
characteristic behaviour of the system.}
\end{center} 
\end{figure}

\begin{figure}[h]
\begin{center}
\includegraphics[width=0.8\textwidth]{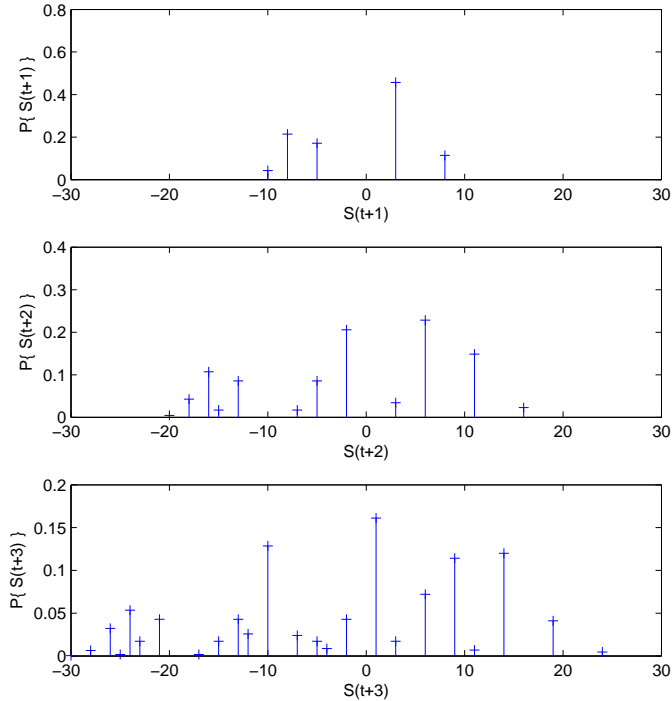}
\caption{ \label{fig:pis} The characteristic behaviour of the system for $U = 1, 2, 3$ time-steps into the future from unknown 
initial state. This is equivalent to looking at the relative frequency of occurrence of changes in output values over 1,2, and 3 
time-step rolling windows}
\end{center} 
\end{figure}

We can now construct the Future-Cast. We can express the evolution of the output in their corresponding states as the 
superposition of the required convolutions:
\begin{eqnarray}\label{eqn:fut1}
\varsigma^{U+1}_A(S)&=& \Upsilon_{A\to A}(\delta)~ \otimes~ \varsigma^{U}_A(S) ~+~ \Upsilon_{B\to A}(\delta)\otimes~ 
\varsigma^{U}_B(S) \nonumber\\
\varsigma^{U+1}_B(S)&=& \Upsilon_{A\to B}(\delta)~ \otimes~ \varsigma^{U}_B(S) ~+~ \Upsilon_{B\to B}(\delta)\otimes~ 
\varsigma^{U}_B(S)
\end{eqnarray}
where $\otimes$ is the discrete convolution operator as defined in footnote~\ref{foot:conv}. Recall that $S(0)$ has been set to 
zero such that we can consider the possible changes in output and the output itself to be identical.  We can write this more 
concisely:
\begin{eqnarray}\label{eqn:fut2}
\underline{\varsigma^{U+1}}&=& \underline{\underline{\Upsilon}}~\underline{\varsigma^{U}}
\end{eqnarray}
Where the element $\Upsilon_{1,1}$ contains the function $\Upsilon_{A\to A}(\delta)$ and the operator $\otimes$ and the element 
$\varsigma^{U+1}_1$ is the distribution $\varsigma^{U+1}_A(S)$.
We note that this matrix of functions and operators is static, so only need computing once. As such we can rewrite 
equation~\ref{eqn:fut2} as
\begin{eqnarray}\label{eqn:fut5}
\underline{\varsigma^{U}}&=& \underline{\underline{\Upsilon}}^U~\underline{\varsigma^{0}}
\end{eqnarray}
such that $\underline{\varsigma^{0}}$ contains the state-wise information of our starting point (time $t$). This is the 
Future-Cast process. For a system starting in state 
\textbf{\emph{A}} and with the start value of the time series of zero, the elements of $\underline{\varsigma^{0}}$ are as shown 
in figure~\ref{fig:start1}.
Applying the Future-Cast process, we can look precisely at the systems potential output at any number of time-steps into
the future. If we wish to consider the process over 10 steps again, we applying as following:
\begin{eqnarray}\label{eqn:fut3}
\underline{\varsigma^{10}}&=& \underline{\underline{\Upsilon}}^{10}~\underline{\varsigma^{0}}
\end{eqnarray}
The resultant distribution of possible outputs (we denote $\Pi^U (S)$) is then just the superposition of the contributions from 
each state. This distribution of the possible outputs at some time into the future is what we call the $Future-Cast$.
\begin{eqnarray}\label{eqn:fut4}
\Pi^{10}&=& \varsigma^{10}_1 ~+~\varsigma^{10}_2
\end{eqnarray}
This leads to distribution shown in figure~\ref{fig:dicefutcast}.
Although the exact output as calculated using the Future-Cast process demonstrated in figure~\ref{fig:dicefutcast} compares well 
with the brute force numerical results of the Monte-Carlo technique in figure~\ref{fig:output2}, it  allows us to perform some 
more interesting analysis without much more work, let alone computational exhaustion.
Consider that we don't know the initial state of the system or that we want to know characteristic properties of the system.
This might include wanting to know what the system does, on average over one time-step increments. We could for example look run 
the system for a very long time and investigate  a histogram of the output movements over single time-steps as shown in 
figure~\ref{fig:hist1}. 
However, we can use the formalism to generate this result exactly. Imagine that model has been run for a long time but we don't 
know which state it is in. Using the probabilities associated with the transitions between the states, we can infer the 
likelihood that the system's internal state is either  \textbf{\emph{A}} or \textbf{\emph{B}}. Let the element $\Gamma^t_1$ 
represent the probability that the system is in state \textbf{\emph{A}} at some time $t$ and $\Gamma^t_2$ that the system is in 
state \textbf{\emph{B}}. We can express these quantities at time $t+1$ by considering the probabilities of going between the 
states:
\begin{eqnarray}\label{eqn:trans}
\Gamma^{t+1}_1 & = & T_{A\to A} \Gamma^{t}_1~+~T_{B\to A} \Gamma^{t}_2\nonumber\\ 
\Gamma^{t+1}_2 & = & T_{A\to B} \Gamma^{t}_2~+~T_{B\to B} \Gamma^{t}_2
\end{eqnarray}
Where $T_{A\to A}$ represents the probability that when the system is in state \textbf{\emph{A}}  it will be in state 
\textbf{\emph{A}} at the next time-step. We can express this more concisely:
\begin{eqnarray}\label{eqn:trans2}
\underline{\Gamma^{t+1}} & = & \underline{\underline{T}}~\underline{\Gamma^{t}}
\end{eqnarray}
Such that $T_{1,1}$ is equivalent to $T_{A\to A}$. This is a Markov Chain. From the nature of our dice, we can trivially 
calculate the elements of  $\underline{\underline{T}}$.  The value for $T_{A\to A}$ is the sum over all elements in 
$\Upsilon_{A\to A}(\delta)$ or more precisely:
\begin{eqnarray}\label{eqn:trans3}
T_{A\to A} = \sum_{\delta = -\infty}^{\infty}\Upsilon_{A\to A}(\delta)
\end{eqnarray}
The Markov Chain transition matrix $\underline{\underline{T}}$ for our system can thus be trivially written:
\begin{eqnarray}\label{eqn:T}
\underline{\underline{T}} = \left(\begin{array}{c c}
0.7 & 0.4 \\
0.3 & 0.6 \end{array}\right)
\end{eqnarray}
The static probabilities of the system being in either of its two states are given by the eigenvector solution to 
equation~\ref{eqn:eig} with eigenvalue $1$. This is equivalent to looking at the relative occurrence of the two states if the 
system were to be run over infinite time. 
\begin{eqnarray}\label{eqn:eig}
\underline{\Gamma} & = & \underline{\underline{T}}~\underline{\Gamma}
\end{eqnarray}
For our system, the static probability associated with being in state \textbf{\emph{A}} is $\frac{4}{7}$ and obviously  
$\frac{3}{7}$ for state \textbf{\emph{B}}. To look at the characteristic properties we are going to construct an initial vector 
similar to $\varsigma^0$ in equation~\ref{eqn:fut1}) for our Future-Cast formalism to act on but which is related to the static 
probabilities contained by the solution to equation~\ref{eqn:eig}. This vector is denoted $\underline{\kappa}$ and its form is 
described in figure~\ref{fig:kappa}.
We use employ $\underline{\kappa}$ in the Future-Cast over one time-step as in equation~\ref{eqn:fut6}:
 \begin{eqnarray}\label{eqn:fut6}
\underline{\varsigma^{1}}&=& \underline{\underline{\Upsilon}}~\underline{\kappa}
\end{eqnarray}
The resulting distribution of possible outputs from superimposing the elements of $\underline{\varsigma}^1$ is the exact 
representation of the one time-step increments ($S(t+1)~-S(t)$) of the system if it were allowed to be run infinitely.
 We call this characteristic distribution $\Pi_{char}^1$. Applying the process a number of times will yield the exact 
distributions $\Pi_{char}^U$ equivalent to looking at all values of $S(t+U)~-S(t)$ which is a rolling window length $U$ over an 
infinite time series as in figure~\ref{fig:pis}. This is also equivalent to running the system forward in time $U$ time-steps 
from unknown initial state, investigating all possible
paths. The Markovian nature of the system means that this is not the same as the convolution of the one time-step characteristic 
Future-Cast convolved with it self $U$ times. 

Clearly the characteristic Future-Cast over one time-step in figure~\ref{fig:pis} compares well with that of 
figure~\ref{fig:hist1}.

\newpage
\section{The Evolution of the Complex Adaptive System}\label{sec:formal}
Here we provide a general formalism applicable to any Complex System which can be mapped
onto a population of $N$ species or `agents' who are repeatedly taking actions in some form of
global `game'. At each time-step each agent makes a (binary) decision
$a_{\mu(t)}$  in response to the global information $\mu(t)$ which may reflect the history of
past global outcomes. This global information is of the form of a bitstring of length $m$. For a
general game, there exists some winning outcome $w(t)$ based on the aggregate action of the
agents. Each agent holds a subset of all possible strategies - by assigning this subset randomly
to each agent, we can mimic the effect of large-scale heterogeneity in the population. In other
words, we have a simple way of generating a potentially diverse ecology of species, some of which may be
similar but others quite different. One can hence investigate a typically-diverse ecology 
whereby all possible species are represented, as opposed to special cases of ecologies which may
themselves generate pathological behaviour due to their lack of diversity. 

\begin{figure}
\begin{center}
\includegraphics[width=0.75\textwidth]{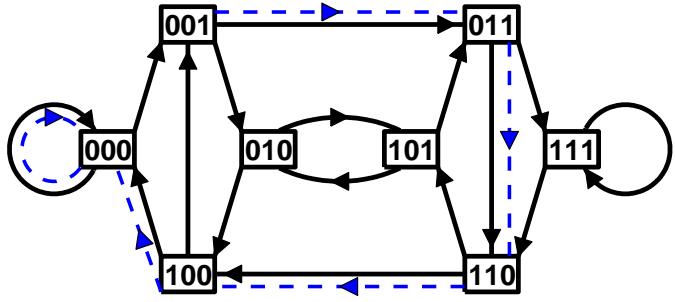}
\caption{\label{fig:debruijn}
A path of time-horizon length $H=5$ (dashed line) superimposed on the
de Bruin graph for $m=3$.  The 8 global outcome states represent the 8 possible bitstrings for the
global information, and correspond to the global outcomes for the past $m=3$ timesteps.}
\end{center} 
\end{figure}

The aggregate action of the population at each time-step $t$
is represented by
$D(t)$, which corresponds to the accumulated decisions of all the agents and hence the (analogue)
output variable of the system at that time-step. The goal of the game, and hence the winning
decision, could be to favour the minority group (MG), the majority group or  indeed any function of
the macroscopic or microscopic variables of the system. The individual agents do not themselves
need to be conscious of the precise nature of the game, or even the algorithm for deciding how the
winning decision is determined. Instead, they just know the global outcome, and hence whether their
own strategies predicted the winning action\footnote{The algorithm used by the `Game-master' to generate the winning decision 
could also
incorporate a stochastic factor.}. The agents then reward the strategies in their possession if
the strategy's predicted action would have been correct if that strategy was implemented.
The global history is then updated according to the winning
decision. It can be expressed in decimal form as follows:
\begin{equation}\label{eqn:formal1}
\mu(t) = \sum_{i=1}^{m}2^{i-1}[w(t-i)+1]\ 
\end{equation}   
The system's dynamics are defined by the rules of the game. 
We will consider here the class of games whereby each agent uses his highest-scoring strategy
at each timestep, and agents only participate if they possess a strategy
with a sufficiently high success rate. [N.B. Both of these assumptions can be relaxed, thereby
modifying the actual game being played]. The following two scenarios might then
arise during the system's evolution:\begin{itemize} 
\item An agent has two (or
         more) strategies which are tied in score and are above the confidence
         level, and the decisions from them differ.
\item The number of
         agents choosing each of the two actions is equal, hence the winning decision is undecided.
\end{itemize} 
We will consider these cases to be
resolved with a fair `coin
         toss', thereby injecting stochasticity or `noise' into the system's dynamical evolution. 
In the first case, each agent will toss his own coin to break the tie, while in the second the
Game-master tosses a single coin.
To reflect the fact that evolving systems will typically be non-stationary, and hence the more
distant past will presumably be perceived as less relevant to the agents, the strategies are 
rewarded as to whether they would have made correct predictions over the last
$H$ time-steps of the game's running. There is no limit on the size of $H$ other than it is finite
and constant. The time-horizon represents a trajectory of length $H$ on the
de Bruijn graph in $\mu(t)$ (history) space\cite{thmg} as shown in figure~\ref{fig:debruijn}. The stochasticity in the game 
means
that for a given time-horizon $H$ and a given strategy allocation in the population, the output of the system is not always 
unique. We will denote the set of all possible outputs from the game at some number of time-steps beyond the time-horizon $H$, 
as the Future-Cast.

It is useful to work in a time-horizon
space $\underline{\Gamma_{t}}$ of dimension $2^{m+H}$. An element $\Gamma_{t}$
corresponds to the last $m+H$ elements of the bitstring of global outcomes (or equivalently,
the winning actions) produced by the game. This dimension is constant in time whereas for a
non-time-horizon game it would grow linearly. For any given time-horizon state, $\Gamma_{t}$, there
exists a unique score vector
$\underline{G(t)}$ which is the set of scores $G_{R}(t)$ for all the strategies which an agent
could possess. As such, for each particular time-horizon state, there exists a unique probability distribution of the aggregate 
action, $D(t)$. This distribution of possible actions when a specified state is reached will necessarily be the same each time 
that state is revisited. Thus, it is
possible to construct a transition matrix (c.f. Markov Chain\cite{thmg}) 
$\underline{\underline{T}}$ of probabilities for the
movements between these time-horizon states such that 
$\underline{P(\Gamma_{t})}$ can be expressed as 
\begin{equation}\label{eqn:formal2}
  \underline{P(\Gamma_{t})} ~= 
  ~\underline{\underline{T}}
  ~\underline{P(\Gamma_{t-1})}\end{equation}
where $\underline{P(\Gamma_{t})}$ is a vector of dimension $2^{m+H}$ containing the probabilities of being in a given state 
$\Gamma$ at time $t$

The transition matrix of probabilities is constant in time and
necessarily sparse. For each state, there are only two possible
winning decisions. The number of
non-zero elements in the matrix is thus $\le2^{(m+H+1)}$.
We can use the transition matrix in an eigenvector-eigenvalue problem to obtain
the stationary state solution of $\underline{P(\Gamma)}
~=\underline{\underline{T}}~\underline{P(\Gamma)} $. This also allows
calculation of some time-averaged macroscopic quantities of the
game \cite{thmg}\footnote{The steady state eigenvector solution is an exact expression equivalent to pre-multiplying the 
probability state vector $\underline{P(\Gamma_t)}$ by $\underline{\underline{T}}^{\infty}$. This effectively results in a 
probability state vector which is time-averaged over an infinite time-interval.}.

To generate the Future-Cast, we want to calculate the
quantities in output space. To do this, we require;
\begin{itemize}\item The probability distribution of $D(t)$ for a given
  time-horizon; \item The corresponding winning decisions, $w(t)$, for
  given $D(t)$; \item An algorithm generating output in terms of $D(t)$.
\end{itemize} 

To implement the Future-Cast, we need to map from the transitions in the state space internal to the
system to the macroscopic observables in the output space (often cumulative excess demand). 
We know that in the transition matrix, the probabilities represent the summation over a
distribution of possible aggregate actions which is binomial in the case where the agents are limited to two possible
decisions. Using the output generating algorithm, we can
construct an `adjacency' matrix $\underline{\underline{\Upsilon}}$ analogous to
the transition matrix $\underline{\underline{T}}$, with the same
dimensions. The elements of $\underline{\underline{\Upsilon}}$, contain probability distribution functions of change in output  
corresponding to the non-zero 
elements of the transition matrix
together with the discrete convolution operator $\otimes$ whose form
depends on that of the output generating algorithm. 

The adjacency matrix of functions and operators can then be applied to a vector, $\underline{\varsigma^{U=0}(S)}$,
containing information about the current state of the game and of the same dimension as $\underline{\Gamma_{t}}$ .  
$\underline{\varsigma^{U=0}(S)}$ not only describes the time-horizon state positionally through its elements but also the 
current value in the output quantity $S$ within that element.
At $U=0$, the state of the system is unique so there is only one non-zero element within $\underline{\varsigma^{U=0}(S)}$. This 
element corresponds to a probability distribution function of the current output value, its position within the vector 
corresponding to the current time-horizon state.
The probability distribution function is necessarily of value unity at the
current value or, for a Future-Cast expressed in terms of change in output from the current value, unity at the
origin. The Future-Cast process for $U$ time-steps beyond the
present state can then be described by
\begin{equation}\label{eqn:formal3}
  \underline{\varsigma^U(S)}~=~ \underline{\underline{\Upsilon}}^{U} \underline{\varsigma^0(S)}
  \end{equation}

The actual Future-Cast, $\Pi(S,U)$, is then computed by superimposing
the elements of the output/time-horizon state vector:
\begin{equation}\label{eqn:formal4}
\Pi^U(S) = \sum_{i=1}^{2^{(m+H)}}\varsigma_i^U(S).\end{equation}
Thus the Future-Cast,  $\Pi^U(S)$, is a probability distribution of the outputs possible at $U$ time-steps in the future. 

As a result of the state dependence of the Markov Chain, $\Pi$ is non-Gaussian. As with the steady-state solution
of the state space transition matrix, we would like to find a `steady-state' equivalent for the output space\footnote{Note that 
we can use
  this framework to generate time-averaged quantities of {\em any} of the macroscopic quantities of the system (e.g total number 
of agents playing) or volatility.} of the
form
\begin{equation}\label{eqn:formal5}
\Pi_{char}^1(S) = \big<\Pi^1(S)\big>_\infty   
\end{equation}
where the one-timestep Future-Cast is time-averaged over an infinitely
long period. 
Fortunately, we have the steady state solutions of $\underline{P(\Gamma)}
~=\underline{\underline{T}}~\underline{P(\Gamma)} $ which are the (static)
probabilities of being in a given time-horizon state at any time. By representing
these probabilities as the appropriate functions, we can construct an
`initial' vector, $\underline{\kappa}$, similar in form to
$\underline{\varsigma(S,0)}$ in equation~\ref{eqn:formal3} but equivalent to the eigenvector solution of the Markov Chain. We 
can then generate the solution of  equation~\ref{eqn:formal5} for
the $characteristic$ Future-Cast, $\Pi_{char}^1$, for a given initial set of
strategies. An element $\kappa_{i}$ is again a probability distribution which is simply the 
point (0 , $P_{i}(\Gamma)$), the static probability of being in the time-horizon state denoted by the elements position, $i$. We 
can then get back to the Future-Cast 
\begin{equation} \label{eqn:formal6}
\Pi_{char}^{1}(S)~=~
  \sum_{i=1}^{2^{(m+H)}}\varsigma_i^1~~~\textrm{where}~~~\underline{\varsigma^1}~=~\underline{\underline{\Upsilon}}~ 
\underline{\kappa}.\end{equation}
We can also generate characteristic Future-Casts for any number of
time-steps, $U$, by pre-multiplying $\underline{\kappa}$ by $\underline{\underline{\Upsilon}}^U$ 
\begin{equation}  \label{eqn:formal7}
\Pi_{char}^{U}(S)~=~
\sum_{i=1}^{2^{(m+H)}}\varsigma_i^U~~~\textrm{where}~~~\underline{\varsigma^U}~=~
\underline{\underline{\Upsilon}}^U~\underline{\kappa}\
\end{equation} 
We note that $\Pi_{char}^{U}$ is not equivalent to the convolution of
$\Pi_{char}^1$ with itself $U$ times and as such is not necessarily Gaussian. 
The characteristic Future-Cast over $U$ time-steps is simply the Future-Cast of length $U$ from all
the $2^{m+H}$ possible initial states where each contribution is given the appropriate weighting
factor. This factor corresponds to the probability of being in that initial state. The characteristic
Future-Cast can also be expressed as
\begin{equation}\label{eqn:formal8}
\Pi_{char}^U (S)~~ =~~ \sum_{\Gamma=1}^{2^{(m+H)}}P(\Gamma)~ \Pi^U(S)\mid\Gamma \end{equation}

where  $\Pi^U(S)\mid\Gamma $ is a normal Future-Cast from an initial time-horizon state $\Gamma$  
and $P(\Gamma)$ is the static probability of being in that state at a given time.

\newpage
\section{The Binary Agent Resource System}\label{sec:BAR}
The general binary framework of the B-A-R (Binary Agent Resource) system was discussed in
section~\ref{sec:formal}. The global outcome of the `game' is represented as a binary digit which favours
either those choosing option
$+1$ or option $-1$ (or equivalently $1$ or $0$, A or B etc.).  
The agents are randomly assigned $s$ strategies at the beginning of
the game. Each strategy comprises an action $a_{\mu(t)}^s$ in response to each of the
$2^{m}$ possible histories
$\mu$, thereby generating a total of $2^{2^{m}}$ strategies in the Full Strategy Space
\footnote{We note that many features of the game can be reproduced using
a Reduced Strategy Space of $2^{m+1}$ strategies, containing strategies which are either
anti-correlated or uncorrelated with each other\cite{A246}. The framework established in the present paper is
general to both the full and reduced strategy spaces, hence the full strategy space will be
adopted here.}. At each turn of the game, the agents employ their most successful
strategy, being the one with the most virtual points. The agents are thus adaptive if $s>1$. 
\begin{figure}
\begin{center}
\includegraphics[width=0.9\textwidth]{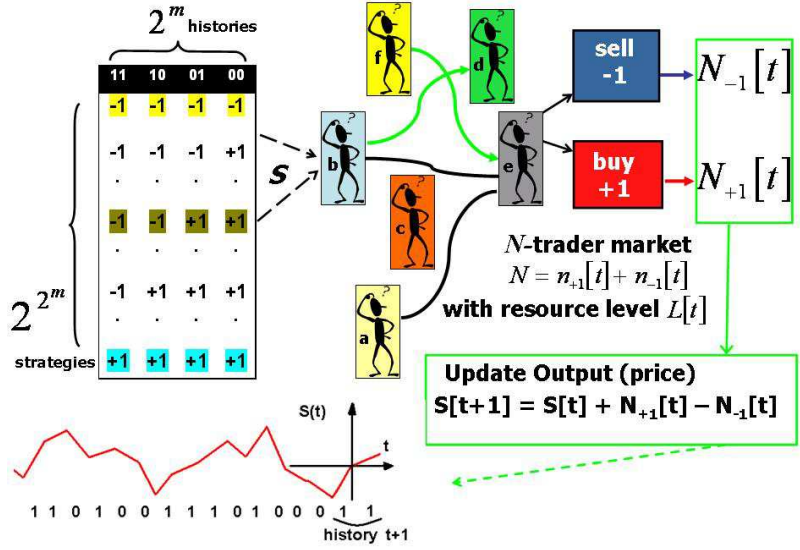}
\caption{\label{fig:BARscheme} Schematic diagram of the Binary Agent Resource (B-A-R) system.}
\end{center}
\end{figure}

We have already extended the B-A-R system by introducing the time-horizon $H$, which determines
the number of past time-steps over which virtual points are
collected for each strategy. We further extend the system  by the introduction of a confidence level.
The agents decide whether to participate or not depending on the success of
their strategies. As such, the number of active agents $N(t)$ is less than or equal to $N_{tot}$ at
any given time-step. This results in a variable number of participants per time-step $V(t)$, and
constitutes a `Grand Canonical' game. The threshold,
$\tau$, denotes the confidence level: each agent will only participate if he has a
strategy with
at least $r$ points where \begin{equation} r = T(2\tau-1).\end{equation}
Agents without an active strategy become temporarily inactive. 

In keeping with typical biological, ecological, social or computational systems, the Game-master
takes into account a finite global resource level when deciding the winning decision at each
time-step. For simplicity, we will here consider the specific case\footnote{We note that $\phi$ itself could be actually be a 
stochastic
function of the known system parameters.} whereby the resource level 
$L(t) =
\phi$$V(t)$ with
$0\le$$\phi$$\le$$1$.
We denote the number of agents choosing action  $+1$ (or equivalently A) as
$N_{+1}(t)$, and those that choose action -1 (or equivalently B) as $N_{-1}(t)$. If
$L(t)-N_{+1}(t)>0$ the winning action is $+1$ and vice-versa. We define the  winning
decision $1$ or $0$ as follows: 
\begin{equation} w(t) = {\rm step}[L(t) - N_{+1}(t)]\end{equation} where we define 
${\rm step}[x]$ to be
\begin{equation} {\rm step}[x] = \left\{\begin{array}{lll} 1 & \textrm{if
}x>0,\\ 0 & \textrm{if }x<0,\\ \textrm{fair coin toss} &
         \textrm{if }x
         = 0.\end{array}
         \right. \end{equation}  
When $x~=~0$, there is no definite winning option since $N_{+1}(t) = N_{-1}(t)$, hence the Game-master
uses a random coin-toss to decide between the two possible outcomes.
We use a binary payoff rule for rewarding strategy scores, 
although more complicated versions can, of course, be used.  
However, we note that non-binary payoffs (e.g. a proportional payoff scheme) will 
decrease the probability of
         tied strategy scores, hence making the system more deterministic. Since we are interested
in seeing the extent to which stochasticity can prevent control, we are instead interested in preserving the presence of such 
stochasticity. The reward function $\chi$ can be written
                  \begin{equation} \chi[N_{+1}(t),L(t)]
         =\left\{\begin{array}{ll} 1 & \textrm{for }
        w(t)=1,\\ -1 & \textrm{for } w(t)=0,\end{array}
         \right. \end{equation} 
namely +1 for predicting the correct action and -1 for predicting the
         incorrect one. For a given strategy, $R$, the virtual points
         score is given by\begin{equation}G_{R}(t) =
         \sum_{i=t-T}^{t-1}a_{R}^{\mu(i)}\chi[N_{+1}(i),L(i)],\end{equation}
         where $a_{R}^{\mu(t)}$ is the response of strategy, $R$, to
         the global information $\mu(t)$ summed over the rolling window
         of width $H$. The global output signal $D(t) = N_{+1}(t)- N_{-1}(t)$ is calculated at
         each iteration to generate an output time series. 
\newpage
\section{Looking at the System's Natural Evolution}\label{sec:natural}
To realize all possible paths within a given game is necessarily
computationally expensive. For a Future-Cast $U$ timesteps beyond the
current game state, there are necessarily $2^{U}$ winning decisions to
be considered. Fortunately, not all winning
decisions are realized by a specific game and the numerical generation
of the Future-Cast can be made reasonably efficient.

Fortunately we can approach the Future-Cast analytically {\em without} having to keep track of
the agents' individual microscopic properties. Instead we group the agents together
via the 
population tensor of rank $s$ given by $\underline{\underline{\Omega}}$, which we will
refer to as the Quenched Disorder Matrix (QDM) \cite{A269}. This matrix is assumed to be
constant over the time-scales of interest, and more typically is fixed at the beginning of the
game. The entry
 $\Omega_{R2,R2,\ldots{}}$ represents the number of agents holding the
 strategies ${R1,R2,\ldots{}}$ such that
 \begin{equation}\sum_{R,R',\ldots{}}\underline{\underline{\Omega}}_{R,R',\ldots{}} = N \end{equation} 
For numerical analysis, it is useful to construct a symmetric version
 of this population tensor, $\underline{\underline{\Psi}}$ . For the case $s=2$, we will let
$\underline{\underline{\Psi}}$ = $\frac{1}{2}(\underline{\underline{\Omega}}$+
$\underline{\underline{\Omega}}^{transpose})$ \cite{PRL82}.

The output variable $D(t)$ can be written in terms of the decided agents
$D_d(t)$ who act in a pre-determined way since they have a unique predicted action from their
strategies, and the undecided agents $D_{ud}(t)$ who require an additional coin-toss in order to
decide which action to take. Hence
\begin{equation}D(t)=D_d(t)+D_{ud}(t).\end{equation} We focus on $s=2$ strategies per agent
although the approach can be generalized. The element $\Psi_{R,R'}$ represents the number of
agents holding both strategy $R$ and $R'$. We can now write $D_d(t)$
as 
\begin{equation} D_d(t) =
 \sum_{R=1}^{Q}a_{R}^{\mu(t)}\mathcal{H}[G_{R}(t)-r]\sum_{R'=1}^{Q}(1+\mbox{sgn}[G_{R}(t)-G_{R'}(t)])\Psi_{R,R'}
 \end{equation}
where $Q$ is the size of the strategy space, $\mathcal{H}$ is the Heaviside function and \mbox{sgn}$[x]$ is defined as 
\begin{equation} \mbox{sgn}[x] =
\left\{\begin{array}{lll} 1 & \textrm{if }x>0,\\ -1 & \textrm{if }x<0,\\ \textrm{0} &
         \textrm{if }x
         = 0.\end{array}
         \right. \end{equation} 
The volume $V(t)$ of active agents  can be 
expressed as
\begin{equation}
V(t)~=~ \sum_{R,R'}\mathcal{H}[G_{R}(t)-r]\big\{sgn[G_{R}(t)-G_{R'}(t)]+\frac{1}{2}\delta[G_{R}(t)-G_{R'}(t)]
\big\}\Psi_{R,R'}
\end{equation}
where $\delta$ is the Dirac delta. The number of undecided agents $N_{ud}(t)$ is given by
\begin{equation} N_{ud}(t)~=~ 
\sum_{R,R'}\mathcal{H}[G_{R}(t)-r]\delta(G_{R}(t)-G_{R'}(t))[1-\delta(a_{R}^{\mu(t)}-a_{R'}^{\mu(t)})]
\Psi_{R,R'}\end{equation}
We note that for $s=2$, because each undecided agent's contribution to $D(t)$ is an integer, hence the
demand of all the undecided agents $D_{ud}(t)$ can be written simply as
\begin{equation} D_{ud}(t)~ \epsilon~
 2 ~\mbox{Bin}\bigg(N_{ud}(t),\frac{1}{2}\bigg)-N_{ud}(t) \end{equation}
where \mbox{Bin}$(n,p)$ is a sample from a binomial distribution of $n$ trials
with probability of success $p$.

For any given time-horizon space-state $\Gamma_{t}$, the score vector $\underline{G(t)}$ (i.e., the
set of scores $G_{R}(t)$ for all the strategies in the QDM) is
unique. Whenever this
state is reached, the quantity $D_d(t)$ will necessarily always be
the same, as will the distribution of $D_{ud}(t)$. We can now construct the  transition matrix
$\underline{\underline{T}}$ giving the probabilities for the
movements between these time-horizon states.
The element $\underline{\underline{T}}_{\Gamma_{t}\mid\Gamma_{t-1}}$
which corresponds to the transition from state $\Gamma_{t-1}$ to
$\Gamma_{t}$, is given for the (generalisable) $s=2$ case by
\begin{eqnarray} 
  \underline{\underline{T}}_{\Gamma_{t}\mid\Gamma_{t-1}} =
  \sum_{x=0}^{N_{ud}}\Bigg\{{}^{N_{ud}}C_x(\frac{1}{2})^{N_{ud}}\delta\bigg[Sgn(D_d+2x-N_{ud}
  +V(1-2\phi))~+{}\nonumber\\(2\mu_t\% 2-1)\bigg]~+
  {}\nonumber\\{}^{N_{ud}}C_x(\frac{1}{2})^{(N_{ud}+1)}\delta\bigg[Sgn(D_d+2x-N_{ud}+V(1-2\phi))~+~0\bigg]
  \Bigg\}\end{eqnarray} where $N_{ud}$, $D_d$ implies $N_{ud}\mid \Gamma_{t-1}$
  and $D_d\mid \Gamma_{t-1}$, $V$ implies $V(t-1)$, $\phi$ sets the resource level as described earlier and $\mu_t \% 2$ is the
  required winning decision to get from state $\Gamma_{t-1}$ to state
  $\Gamma_{t}$.
We use the transition matrix in the eigenvector-eigenvalue problem to obtain
the stationary state solution of $\underline{P(\Gamma)}
~=\underline{\underline{T}}~\underline{P(\Gamma)} $.
The probabilities in the transition matrix represent the summation over a distribution which
is binomial in the $s=2$ case. These distributions are all
calculated from the QDM which is fixed from the outset.
To transfer to output-space, we require an output generating algorithm. Here we use the equation
\begin{equation}S(t+1) = S(t)+D(t)\end{equation} 
hence the 
output value $S(t)$ represents the cumulative value of $D(t)$, while the
increment $S(t+1)-S(t)$ is simply $D(t)$. Again, we use the discrete convolution operator
$\otimes$ defined as \begin{equation}
 (f\otimes g)\mid_{i} ~= \sum_{j=-\infty}^{\infty} 
f(i-j)\times g(j).\end{equation}
The formalism could be extended for general output algorithms using
differently defined convolution operators. 

An element in the adjacency matrix for the $s=2$ case can then be
expressed as
\begin{eqnarray}
\underline{\underline{\Upsilon}}_{\Gamma_{t}\mid\Gamma_{t-1}} =
  \Bigg\{\sum_{x=0}^{N_{ud}}\Bigg((D_d+2x-N_{ud}),~~~~~~~~~~~~~~~~~~~~~~~{}\nonumber\\ 
{}^{N_{ud}}C_x(\frac{1}{2})^{N_{ud}}\delta\bigg[Sgn(D_d+2x-N_{ud}+V(1-2\phi))+{}
(2\mu_t \% 2-1)\bigg]+{}\nonumber\\ 
{}^{N_{ud}}C_x(\frac{1}{2})^{(N_{ud}+1)}\delta\bigg[Sgn(D_d+2x-N_{ud}+V(1-2\phi))+0\bigg]
   \Bigg)\Bigg\}\otimes~~ \end{eqnarray} where $N_{ud}$, $D_d$ again
implies $N_{ud}\mid \Gamma_{t-1}$ and $D_d \mid\Gamma_{t-1}$, $V$ implies
$V(t-1)$, and $\mu_t\% 2$ is the winning decision necessary to move
between the required states. 
The Future-Cast and characteristic Future-Casts ($\Pi^U(S)$, $\Pi_{char}^U $) $U$ time-steps into the future
  can then be computed for a given initial quenched
disorder matrix (QDM).

\begin{figure}[!h]
\begin{center}
\def\capfrac{1}
\includegraphics[width=0.75\textwidth]{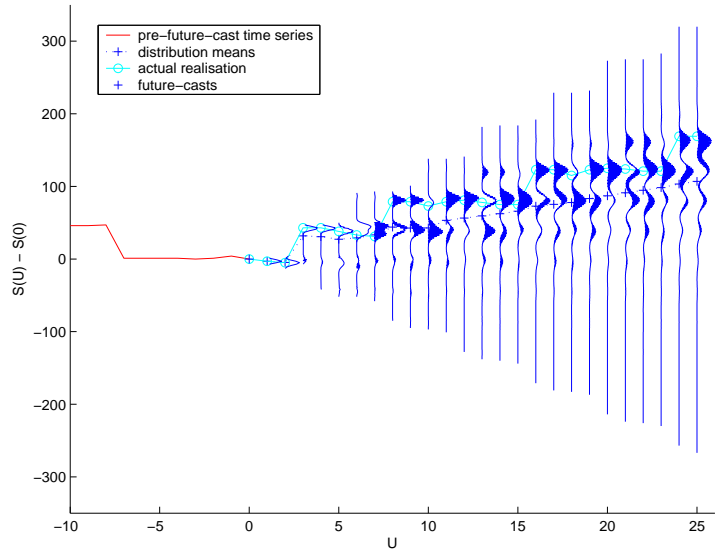}
\caption{\label{fig:res1} The (un-normalized) evolution of a Future-Cast for the given ${\underline{\underline{\Omega}}}$, game 
parameters and initial state $\Gamma$. The figure shows the last 10 time-steps prior to the Future-Cast, the means of the 
distributions within the Future-Cast itself, and also an actual realization of the game run forward in time.}
\end{center}
\end{figure}

We now consider an example to illustrate the implementation. In particular, we provide the explicit
solution of a Future-Cast in the regime of small $m$ and $H$, given the
randomly chosen quenched disorder matrix\begin{tiny}
\begin{equation}
\underline{\underline{\Omega}} =  \left(\begin{array}{cccccccccccccccc}
0 & 0 & 1 & 0 & 0 & 1 & 0 & 1 & 0 & 0 & 1 & 0 & 0 & 0 & 0 & 0\\
0 & 0 & 0 & 1 & 0 & 0 & 1 & 0 & 2 & 0 & 0 & 1 & 0 & 1 & 0 & 0\\
0 & 1 & 1 & 0 & 1 & 0 & 1 & 0 & 0 & 0 & 0 & 2 & 1 & 0 & 0 & 1\\
0 & 1 & 0 & 0 & 2 & 0 & 1 & 0 & 0 & 0 & 0 & 0 & 0 & 0 & 0 & 2\\
1 & 0 & 0 & 0 & 0 & 1 & 0 & 0 & 0 & 1 & 1 & 1 & 1 & 1 & 0 & 1\\
1 & 0 & 0 & 0 & 1 & 1 & 0 & 0 & 1 & 0 & 0 & 0 & 0 & 1 & 2 & 0\\
0 & 1 & 1 & 0 & 0 & 0 & 0 & 0 & 0 & 1 & 0 & 2 & 1 & 0 & 0 & 1\\
0 & 1 & 0 & 0 & 1 & 1 & 0 & 0 & 0 & 1 & 0 & 1 & 0 & 1 & 0 & 1\\
2 & 1 & 0 & 0 & 0 & 1 & 1 & 1 & 0 & 0 & 0 & 0 & 0 & 0 & 0 & 0\\
0 & 0 & 0 & 0 & 0 & 0 & 1 & 0 & 0 & 3 & 1 & 1 & 0 & 1 & 0 & 1\\
0 & 2 & 0 & 0 & 1 & 0 & 0 & 1 & 1 & 3 & 2 & 0 & 0 & 0 & 1 & 1\\
0 & 0 & 2 & 0 & 0 & 0 & 0 & 0 & 0 & 0 & 0 & 0 & 0 & 0 & 0 & 0\\
1 & 0 & 1 & 1 & 0 & 0 & 0 & 0 & 1 & 0 & 0 & 0 & 0 & 0 & 0 & 0\\
0 & 0 & 0 & 1 & 0 & 0 & 0 & 2 & 1 & 0 & 1 & 0 & 0 & 0 & 1 & 0\\
0 & 2 & 0 & 1 & 0 & 0 & 0 & 0 & 0 & 0 & 1 & 0 & 0 & 1 & 0 & 0\\
0 & 0 & 0 & 1 & 0 & 0 & 1 & 0 & 0 & 0 & 0 & 0 & 1 & 0 & 1 &
1\end{array} \right).\end{equation}\end{tiny}
We consider the full strategy space and the the following game parameters: 
\begin{center}\begin{tabular}{c c c}
Number of agents & $N_{tot}$ & 101\\
Memory size & $m$ & 2\\
Strategies per agent & $s$ & 2\\
Resource level & $\phi$& 0.5\\
Time horizon & $H$ & 2\\
Threshold & $\tau$ & 0.51 \end{tabular}\end{center}
The dimension of the transition matrix is thus
$2^{H+m} = 16$.\begin{tiny}\begin{equation}
\underline{\underline{T}} =  \left(\begin{array}{cccccccccccccccc}
0      &0&0&0&0&0&0&0 & 0 &0&0&0&0&0&0&0\\
1      &0&0&0&0&0&0&0 & 1 &0&0&0&0&0&0&0\\
0& 0.5   &0&0&0&0&0&0&0&  0.0625  &0&0&0&0&0&0\\
0& 0.5   &0&0&0&0&0&0&0&  0.9375  &0&0&0&0&0&0\\
0&0& 1    &0&0&0&0&0&0&0&  1  &0&0&0&0&0\\
0&0& 0    &0&0&0&0&0&0&0&  0  &0&0&0&0&0\\
0&0&0& 0.1875     &0&0&0&0&0&0&0&  1 &0&0&0&0\\
0&0&0& 0.8125     &0&0&0&0&0&0&0&  0  &0&0&0&0\\
0&0&0&0& 0.1875    &0&0&0&0&0&0&0&  0.1094  &0&0&0\\
0&0&0&0& 0.8125     &0&0&0&0&0&0&0&  0.8906  &0&0&0\\
0&0&0&0&0& 0     &0&0&0&0&0&0&0&  0.0312  &0&0\\
0&0&0&0&0& 1  &0&0&0&0&0&0&0&  0.9688  &0&0\\
0&0&0&0&0&0& 0.75  &0&0&0&0&0&0&0&  0.5    &0\\
0&0&0&0&0&0& 0.25  &0&0&0&0&0&0&0&  0.5    &0\\
0&0&0&0&0&0&0& 1     &0&0&0&0&0&0&0&     1\\
0&0&0&0&0&0&0& 0     &0&0&0&0&0&0&0&     0
\end{array} \right).\end{equation}\end{tiny}
 Each non-zero element in the transition matrix
corresponds to a probability function in the output space in the
Future-Cast operator matrix. Consider that the initial state is $\Gamma = 10$ i.e. the
last 4 bits are $\{1001\}$ (obtained from running the game prior to the Future-Casting process).
The initial probability state vector is the point (0,1) in
the element of the vector $\underline{\varsigma}$ corresponding to time-horizon state $10$.
We can then generate the Future-Cast for given $U$ (shown in figure~\ref{fig:res1}).

\begin{figure}[!h]
\includegraphics[width=0.45\textwidth]{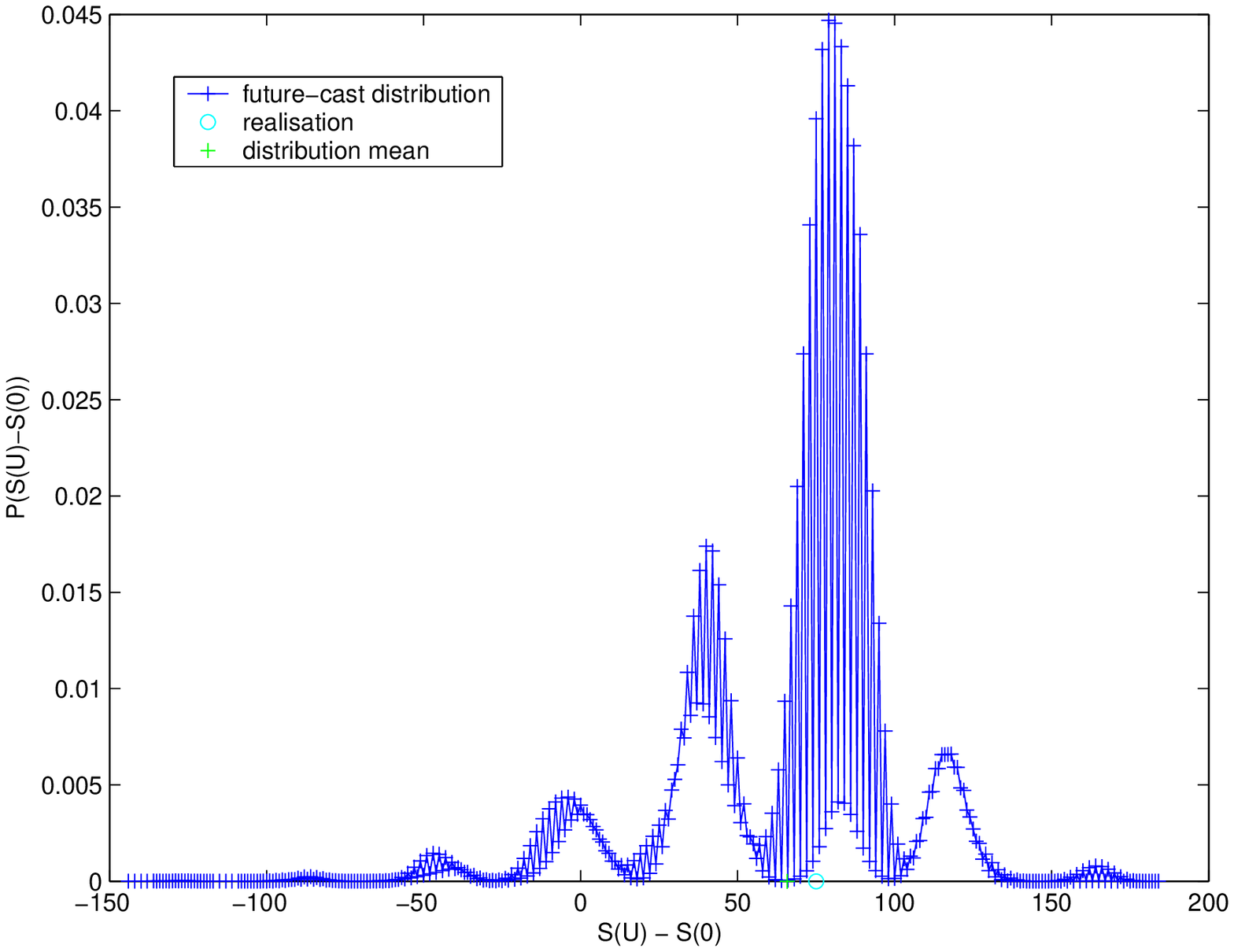}
\includegraphics[width=0.45\textwidth]{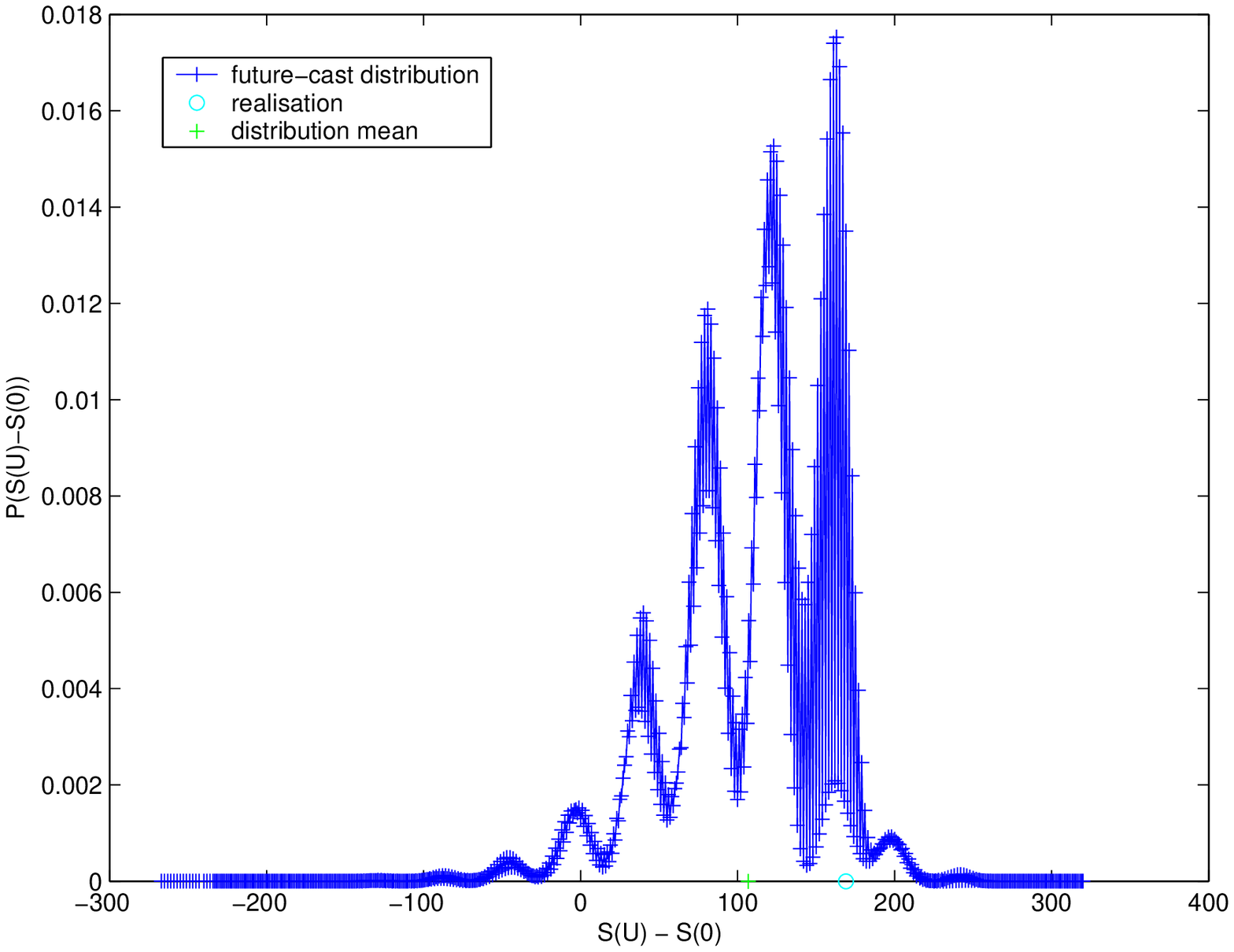}
\caption{\label{fig:res2} The probability distribution function at $U~ =~ 15, 25 $ time-steps
beyond the present state.}
\end{figure}

Clearly the probability function in output space becomes smoother as
$U$ becomes larger and the number of successive convolutions increases, as highlighted by the probability distribution functions 
at $U = 15 $ and $U = 25$ (figure~\ref{fig:res2}).

We note the non-Gaussian form of the probability distribution for the Future-Casts, emphasising the
fact that such a Future-Cast approach is essential for understanding the system's
evolution. An assumption of rapid diffusion toward a Gaussian distribution, and hence the
future spread in paths increasing as the square-root of time, would clearly be unreliable.

\section{Online Evolution Management via `Soft' Control}\label{sec:evolution}
\begin{figure}[!h]
\begin{center}
\includegraphics[width=0.75\textwidth]{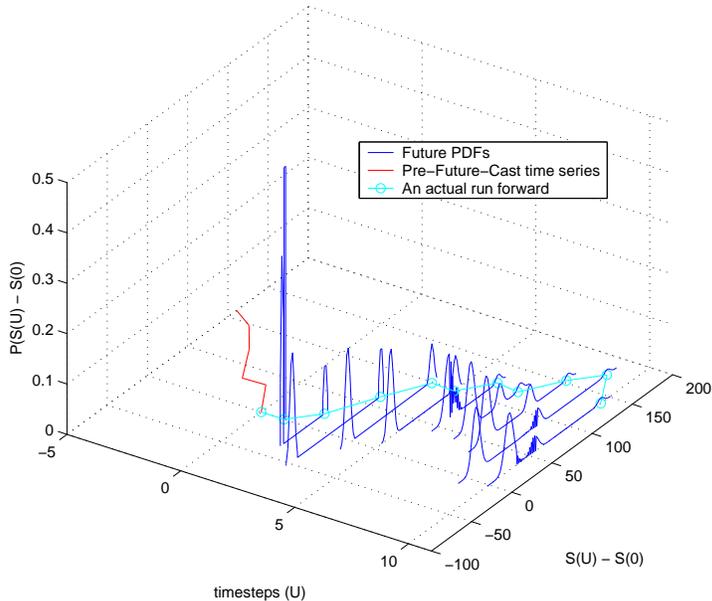}
\caption{\label{fig:thesis3D} Evolution of $\Pi^{U}(S)$ for a typical quenched disorder matrix 
${\underline{\underline{\Omega}}}$.}
\end{center}
\end{figure}

\begin{figure}[!h]
\begin{center}
\includegraphics[width=0.75\textwidth]{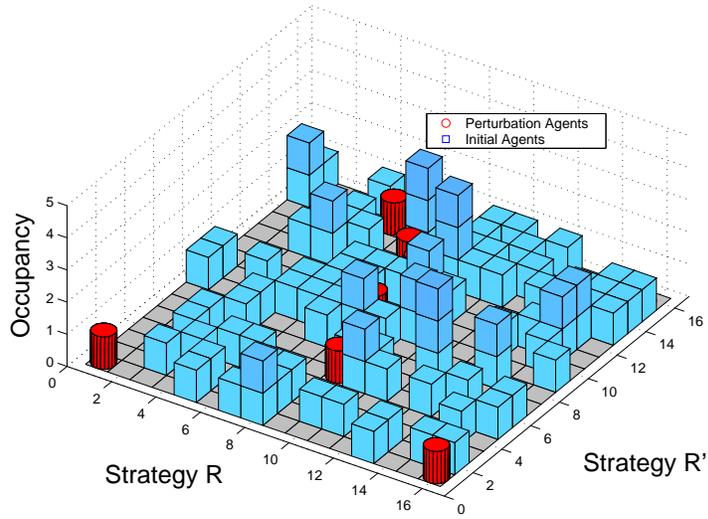}
\caption{\label{fig:figpet}The initial and resulting quenched disorder matrices (QDM), shown in schematic
form. The x-y axes are the strategy labels for the two strategies. The absence of a symbol denotes
an empty bin (i.e. no agent holding that particular pair of strategies).}

\end{center}
\end{figure}

For less simple parameters, the matrix dimension required for the Future-Cast process become very
large very quickly. To generate a Future-Cast appropriate to larger  parameters
e.g. $m =3$, $H = 10$, it is still however possible to carry out the calculations
numerically quite easily. As an example, we generate a random
$\underline{\underline{\Omega}}$ (the form of which is given in figure~\ref{fig:figpet}) and initial time-horizon
appropriate to these parameters.
This time-horizon is obtained by allowing the system to run
prior to the Future-Cast.  For visual representation reasons, the Reduced Strategy Space\cite{A269} is employed. The other game 
parameters are as previously stated. The game is then
instructed to run down every possible winning decision path exhaustively. The spread of output  at
each step along each path is then convolved with the next spread such that a Future-Cast is built
up along each path. Fortunately, not all paths are realized at every time-step since the
stochasticity in the winning-decision/state-space results from the condition $N_{ud}\ge D_d $. The
Future-Cast as a function of
$U$ and $S$, can thus
be built up for a randomly chosen initial quenched disorder matrix (QDM) (\ref{fig:thesis3D}).

\begin{figure}[!h]
\begin{center}
\def\capfrac{1}
\includegraphics[width=0.75\textwidth]{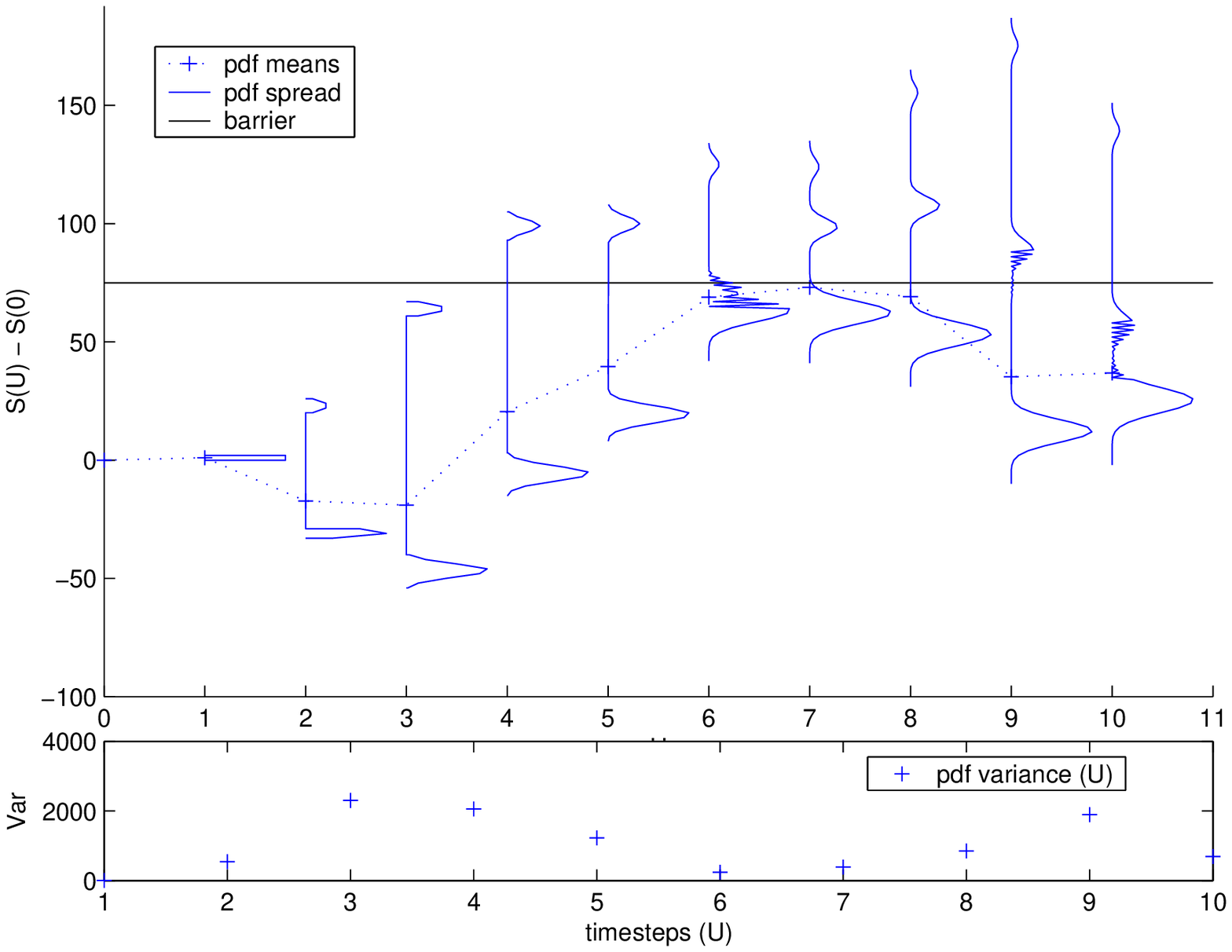}
\caption{\label{fig:figbarrier1} The evolution of the Future-Casts, and the barrier to
be avoided. For simplicity the barrier is chosen to correspond to a fixed $S(t)$ value of 110,
although there is no reason that it couldn't be made time-dependent. Superimposed on the
(un-normalised) distributions, are the means of the Future-Casts, while their variances
are shown below.}
\end{center}
\end{figure}

We now wish to consider the situation where it is required that the system 
should not behave in a certain manner. For example, it may be desirable that it avoid entering a
certain regime characterised by a given value of
$S(t)$. Specifically, we consider the case where there is a barrier in the output space that
the game should avoid, as shown in figure~\ref{fig:figbarrier1}. 

The evolution of the spread (i.e. standard deviation) of the distributions in time, confirms
the non-Gaussian nature of the system's evolution -- we note that this spread can even decrease
with time\footnote{This feature can be understood by appreciating the multi-peaked nature of the
distributions in question. The peaks correspond to differing paths travelled in the
Future-Cast, the final distribution being a superposition of these. If these individual path
distributions mean-revert, the spread of the actual Future-Cast can decrease over short
time-scales.}.
In the knowledge that this barrier will be breached by this system, 
we therefore perturb the quenched disorder at $U = 0$. This perturbation corresponds in physical terms to an
adjustment of the composition of the agent population. This could be achieved by `re-wiring' or
`reprogramming' individual agents in a situation in which the agents were accessible objects, or
introducing some form of communication channel, or even a more `evolutionary' approach whereby a
small subset of species are removed from the population and a new subset added in to replace them.
Interestingly we note that this `evolutionary' mechanism need neither be completely deterministic
(i.e. knowing exactly how the form of the QDM changes) nor completely random (i.e. a random
perturbation to the QDM). In this sense, it seems tantalisingly close to some modern ideas of
biological evolution, whereby there is some purpose mixed with some randomness. 

\begin{figure}[!h]
\begin{center}
\includegraphics[width=0.75\textwidth]{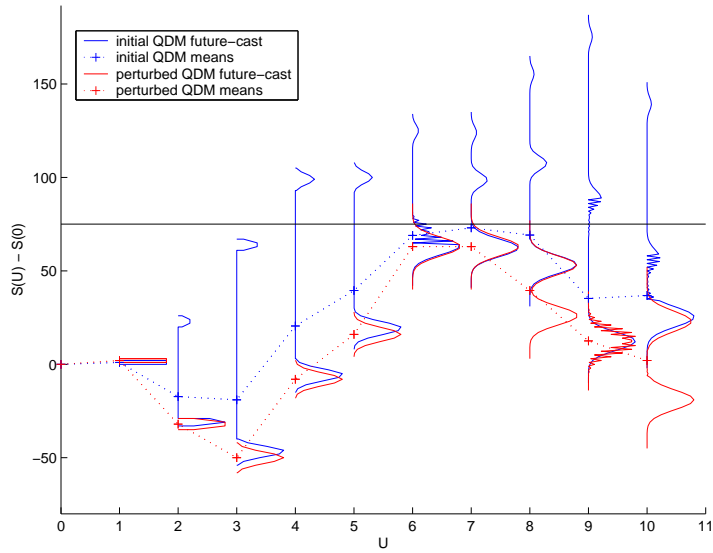}
\caption{\label{fig:figbarrier2}The evolution as a result of the microscopic perturbation to the population's
composition (i.e. the QDM).}
\end{center}
\end{figure}

Figure~\ref{fig:figbarrier2} shows the impact of this relatively minor microscopic perturbation on the
Future-Cast and global output of the system. In particular, the
system has been steered away from the potentially harmful barrier into `safer' territory.

This set of outputs is specific to the initial state of the system. More typically, we may not
know this initial state. Fortunately, we can make use of the characteristic Future-Casts to make
some kind of quantitative assessment of the robustness of the quenched disorder perturbation in
avoiding the barrier, since this procedure provides a picture of the range of possible
future scenarios.

\begin{figure}[!h]
\begin{center}
\includegraphics[width=0.75\textwidth]{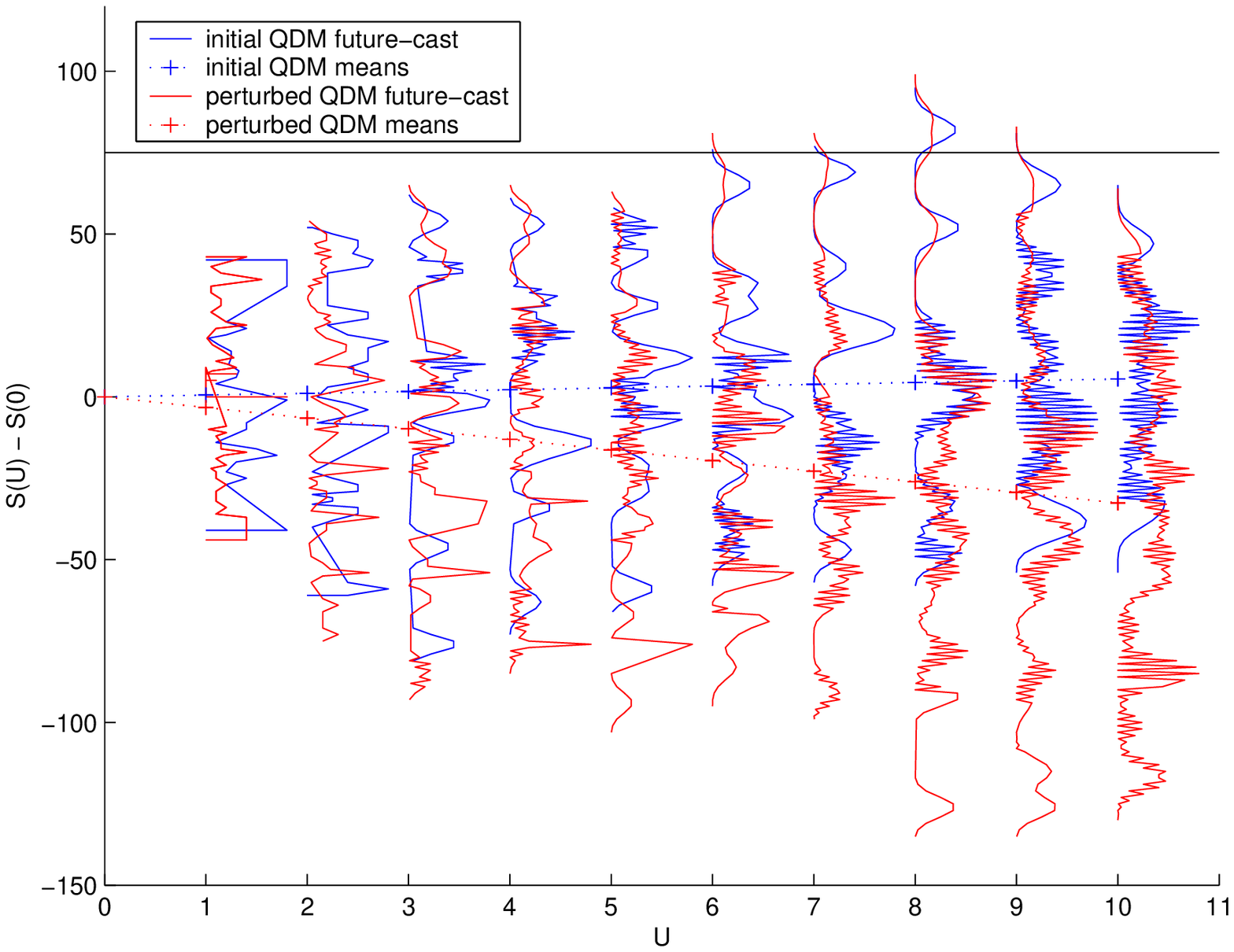}
\caption{\label{fig:charev} The characteristic evolution of the initial and perturbed QDMs.}
\end{center}
\end{figure} 
This evolution of the characteristic Future-Casts, for both the initial
and perturbed quenched disorder matrices, is shown in figure~\ref{fig:charev}
A quantitative evaluation of the robustness of this barrier avoidance could then be 
calculated using traditional techniques of risk analysis, based on knowledge of the distribution
functions and/or their low-order moments.

\section{A Simplified Implementation of the Future-Cast Formalism}\label{sec:reduction}
We introduced the Future-Cast formalism to map from the internal state space of a complex system
to the observable output space. Although the formalism exactly generates the probability distributions of the subsequent output 
from the system, it's implementation is far from trivial. This involves keeping track of numerous distributions and performing 
appropriate convolutions between them. Often, however it is only the lowest order moments which are of immediate concern to the 
system designer. Here, we show how this information can be 
generated without the computational exhaustion previously required. We demonstrate this procedure for a very simple two state 
system, although the formalism is general to a system of any number of states, governed by a Markov chain.

\begin{figure}[h]
\begin{center}
\includegraphics[width=0.75\textwidth]{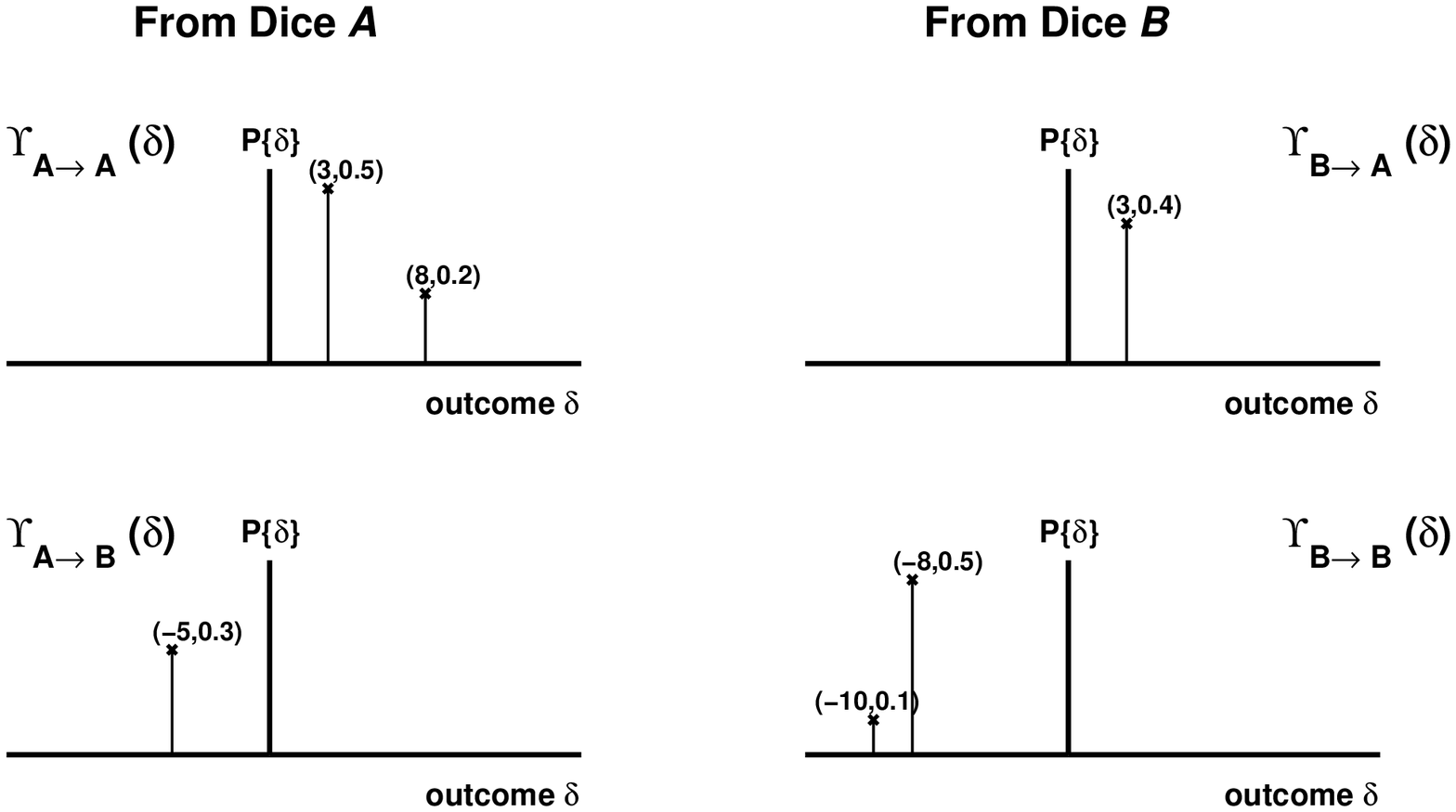}
\caption{ \label{fig:reddicestates2} The state transition distributions as prescribed by our dice.}
\end{center} 
\end{figure}

Recall the toy model comprising two dice of section~\ref{sec:Dice} . We previously broke down the possible outputs of each 
according to the state transition as shown in figure~\ref{fig:reddicestates2}.
These distributions were used to construct the matrix $\underline{\underline\Upsilon}$ to form the Future-Cast process as 
denoted in equation~\ref{eqn:fut2}. This acted on vector $\underline\varsigma^U$
to generate  $\underline\varsigma^{U+1}$. The elements of these vectors contain the partial distribution of outputs which are in 
the state denoted by the element number at that particular time, so for the two dice model, $\varsigma_{1}^U(S)$ contains the 
distribution of output values at time $t+U$ (or $U$ time-steps beyond the present) which correspond to the system being in state 
\textbf{\emph{A}} and $\varsigma_{2}^U(S)$ contains those for state  \textbf{\emph{B}}. 
To reduce the calculation process, we will consider only the moments of each of these individual elements about zero. As such we 
construct a vector, $^{n}\underline{x}_U$, which takes the form:

\begin{eqnarray}
^{n}\underline{x}_U &=& \left( \begin{array}{c}
\sum_{S = -\infty}^{\infty}\varsigma_{1}^U (S) S^n \\
\sum_{S = -\infty}^{\infty}\varsigma_{2}^U(S) S^n 
\end{array}
\right)
\end{eqnarray}

The elements are just the $n$th moments about zero of the partial distributions within the appropriate state. For $n = 0$ this 
vector merely represents the probabilities of being in either state at some time $t+U$.  We note that for the $n = 0$ case, 
$^{0}\underline{x}_{U+1}  =\underline{\underline{T}}~  ^{0}\underline{x}_{U} $ where $\underline{\underline{T}}$ is the Markov 
Chain transition matrix as in equation~\ref{eqn:trans2}. We also note that the transition matrix $\underline{\underline{T}} =~ 
^{0}\underline{\underline{X}} $ where  
we define the (static) matrix  $^{n}\underline{\underline{X}}$ in a similar fashion using the partial distributions  (described 
in figure~\ref{fig:reddicestates2}) to be

\begin{equation}
^{n}\underline{\underline{X}}~=~
\left( \begin{array}{c c}
\sum_{\delta = -\infty}^{\infty}\Upsilon_{A\to A}(\delta) \delta^n  & \sum_{\delta = -\infty}^{\infty}\Upsilon_{B\to A}(\delta) 
\delta^n\\
\sum_{\delta = -\infty}^{\infty}\Upsilon_{A\to B}(\delta) \delta^n  & \sum_{\delta = -\infty}^{\infty}\Upsilon_{B\to B}(\delta) 
\delta^n
\end{array}\right) 
\end{equation}

Again, this contains the moments (about zero) of the partial distributions corresponding to the transitions between states. The 
evolution of $^{0}\underline{x} $ , the state-wise probabilities with time is trivial as described above. For higher orders, we 
must consider the effects of superposition and convolution on their values. We know that the for the superposition of two 
partial distributions, the resulting moments (any order) about zero will be just the sum of the moments of the individual 
distributions, it is just a summation. The effects of convolution, however must be considered more carefully. The elements of 
our vector $^{1}\underline{x}_{U} $  are the first order moments of the values associated with  either of the two states at time 
$t+U$. The first element of which corresponds to those values  of output in state  \textbf{\emph{A}} at time $t+U$. Consider 
that element one step later,  $^{1}{x}_{1,U+1} $. This can be written as the superposition of the two required convolutions.

\begin{eqnarray}
\sum_S~\varsigma_{1}^{U+1}~ S & = &
\sum_{\delta}\sum_{S}~\varsigma_{1}^{U}~\Upsilon_{A\to A}~(S~+~\delta)~+~
 \sum_{\delta}~\sum_{S}~\varsigma_{2}^{U}~\Upsilon_{B\to A}~(S~+~\delta) \nonumber\\
 ^{1}{x}_{1,U+1}& = & ^{0}x_{1,U}~^{1}X_{1,1}~ +~ ^{0}X_{1,1}~^{1}x_{1,U}~+~\nonumber\\
{}&{}&~~~~~~~~^{0}x_{2,U}~^{1}X_{1,2}~ + ~^{0}X_{1,2}~^{1}x_{2,U}
\end{eqnarray}
In simpler terms,
\begin{equation}
^{1}\underline{x}_{U+1} = ~~^{0}\underline{\underline{X}}~^{1}\underline{x}_U~+~
^{1}\underline{\underline{X}}~^{0}\underline{x}_U
\end{equation}
The $(S~+~\delta)$ term in the expression relates to the nature of series generating algorithm, $S_{t+1} = S_t +  \delta_t $. If 
the series updating algorithm were altered, this would have to be reflected in this convolution. 

The overall output of the system is the superposition of the contributions in each state. As such, the  resulting first moment 
about zero (the mean) for the overall output at $U$ time-steps into the future is simply 
$^{1}\underline{x}_{U+1}\cdot\underline{1}$ where $\underline{1}$ is a vector containing all ones and $\cdot$ is the familiar 
dot product.

The other moments about zero can be obtained similarly.
\begin{eqnarray}
^{0}\underline{x}_{U+1}	&=&	~^{0}\underline{\underline{X}}~~^{0}\underline{x}_U	\nonumber\\
^{1}\underline{x}_{U+1}	&=&~^{0}\underline{\underline{X}}~~^{1}\underline{x}_U~+~
~^{1}\underline{\underline{X}}~^{0}\underline{x}_U\nonumber\\
^{2}\underline{x}_{U+1}	&=&~^{0}\underline{\underline{X}}~~^{2}\underline{x}_U~+~
2~^{1}\underline{\underline{X}}~^{1}\underline{x}_U~+~
~^{2}\underline{\underline{X}}~^{0}\underline{x}_U\nonumber\\
^{3}\underline{x}_{U+1}	&=&~^{0}\underline{\underline{X}}~~^{3}\underline{x}_U~+~
3~^{1}\underline{\underline{X}}~^{2}\underline{x}_U~+~
3~^{2}\underline{\underline{X}}~^{1}\underline{x}_U~+~
~^{3}\underline{\underline{X}}~^{0}\underline{x}_U~\nonumber\\
\vdots &{}&{}\vdots
\end{eqnarray}

More generally
\begin{equation}
^{n}\underline{x}_{U+1}	~ =~ \sum_{\gamma=0}^{n} ~^{n}C_\gamma ~^{\gamma}\underline{\underline{X}}~~~^{n-\gamma}\underline{x}_U
\end{equation}
where $^{n}C_\gamma$ is the conventional $choose$ function.

To calculate time-averaged properties of the system, for example the one-time-step mean or variance, we set the initial vectors 
such that
\begin{equation}
^{0}\underline{x}_{0} =~^{0}\underline{\underline{X}}~ ^{0}\underline{x}_{0}
\end{equation}

and $^{\beta}\underline{x}_{0} =  \underline{0}$ for $\beta>0$. The moments about zero can then be used to calculate the moments 
about the mean.
The mean of the one time-step increments in output averaged over an infinite run will then be 
$^{1}\underline{x}_{1}\cdot\underline{1}$ and $\sigma^2$ will be 
\begin{equation}
\sigma^2~=~       ~^{2}\underline{x}_{1}\cdot\underline{1}  -                   (~^{1}\underline{x}_{1}\cdot\underline{1})^2
\end{equation}
These can be calculated for any size rolling window. The mean of all $U$-step increments, $S(t+U)-S(t))$ or conversely the mean 
of the Future-Cast $U$  steps into the future from unknown current state is simply 
$^{1}\underline{x}_{U}\cdot\underline{1}$and $\sigma^2_U$ will be 
\begin{equation}
\sigma^2_U~=~       ~^{2}\underline{x}_{U}\cdot\underline{1}  -                   (~^{1}\underline{x}_{U}\cdot\underline{1})^2
\end{equation}
again with initial vectors calculated from $^{0}\underline{x}_{0} =~^{0}\underline{\underline{X}}~ ^{0}\underline{x}_{0}$ and 
$^{\beta}\underline{x}_{0} =  \underline{0}$ for $\beta>0$.
Examining this explicitly for our two dice model, the initial vectors are:
\begin{eqnarray}
^{0}\underline{x}_0 &=& \left( \begin{array}{c}
\frac{4}{7}\\
\frac{3}{7}
\end{array}\right)\nonumber\\
^{1}\underline{x}_0 &=& \left( \begin{array}{c}
0\\
0
\end{array}\right)\nonumber\\
^{2}\underline{x}_0 &=& \left( \begin{array}{c}
0\\
0
\end{array}\right)\nonumber\\
\end{eqnarray}
and the (static) matrices are:
\begin{eqnarray}
^{0}\underline{\underline{X}}&=&
\left( \begin{array}{c c}
 0.7 & 0.4 \\
 0.3 & 0.6
\end{array}\right) \nonumber\\
^{1}\underline{\underline{X}}&=&
\left( \begin{array}{c c}
    3.1   & 1.2\\
   -1.5  & -5.0
\end{array}\right) \nonumber\\
^{2}\underline{\underline{X}}&=&
\left( \begin{array}{c c}
   17.3 &   3.6\\
    7.5&  42.
\end{array}\right) \nonumber\\
\end{eqnarray}

\begin{figure}[h]
\begin{center}
\includegraphics[width=0.75\textwidth]{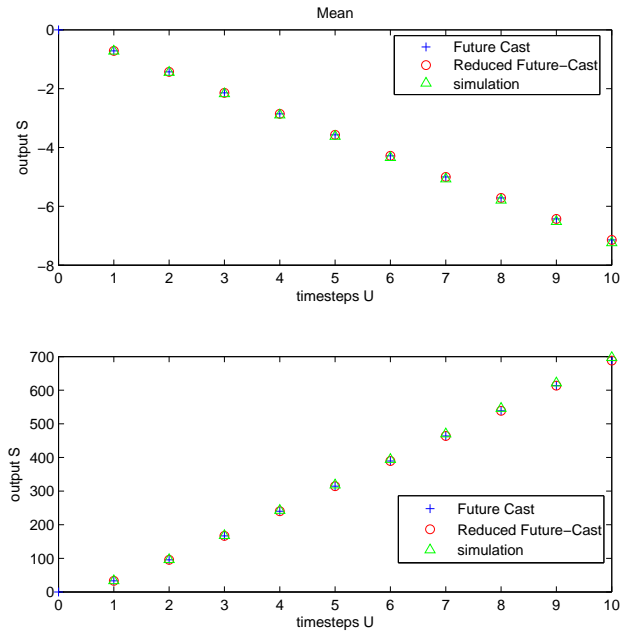}
\caption{ \label{fig:compare} The means and variances of the characteristic distributions $\Pi^1_{char}$ to $\Pi^{10}_{char}$ as 
compared to a numerical evaluation and the reduced Future-Cast.}
\end{center} 
\end{figure}

These are all we require to calculate the means and variances for our system's potential output at any time in the future.
To check that all is well, we employ the Future-Cast to generate the possible future distributions of output up till 10 
time-steps, $\Pi^1_{char}$ to $\Pi^{10}_{char}$. The means and variances of these are compared to the reduced Future-Cast 
formalism and also a numerical simulation. This is a single run of the game over 100000 time-steps. The means and variances are 
then measured over rolling windows of between 1 and 10 time-steps in length. The comparison is shown in 
figure~\ref{fig:compare}.
Fortunately they all concur. The Reduced Future-Cast formalism and the moments about either the mean or zero from the 
distributions generated by the Future-Cast formalism are identical. Clearly numerical simulations require progressively longer 
run times to investigate the properties of distributions further into the future, where the total number of possible paths gets 
large.

\section{Discussion}\label{sec:conclusion}
We have presented an analytical formalism for the calculation of the
probabilities of outputs from the B-A-R system at a number of
time-steps beyond the present state. The construction of the
(static) Future-Cast operator matrix allows the
evolution of the systems output, and other macroscopic quantities of
the system, to be studied without the
need to follow the microscopic details of each agent or species. We have demonstrated the technique
to investigate the macroscopic effects of population perturbations but it could also be used to 
explore the effects of exogeneous noise or even news in the context of financial markets.
We have concentrated on single realisations of the quenched
disorder matrix, since this is appropriate to the 
behaviour and design of a particular realization of a system in practice.
An example could be a financial market model based on the B-A-R system whose derivatives could be analysed quantitatively using 
expectation values generated with the Future-Casts.
We have also shown that through the normalised eigenvector solution of the
Markov Chain transition matrix, we can use the Future-Cast operator
matrix to generate a characteristic probability function for a given game over a given time
period.  The formalism is general to any time-horizon game and could,
for example, be used to analyse  
systems (games) where a level of communication between the agents is permitted, or even linked
systems (i.e. linked games or `markets'). In the context of linked systems, it will then be interesting to
pursue the question as to when adding one `safe' complex system to another `safe' complex system,
results in an `unsafe' complex system. Or thinking more optimistically, when can we put together
two or more `unsafe' systems and get a `safe' one?

We have also presented a simplified and altogether more usable interpretation of the Future-Cast formalism for tracking the 
evolution of the output variable from a complex system whose internal states can be described as a Markov process. We have 
illustrated the application of the results for an example case both for the evolution from a known state or when the present 
state is unknown, to give characteristic information about the output series generated by such a system. The formalism is 
generalizable to Markov Chains whose state transitions are not limited to just two possibilities and also to systems whose 
mapping from state transitions to output-space are governed by continuous probability distributions.

Future work will focus on the `reverse problem' of the broad-brush design of multi-agent systems which behave in some particular desired way -- or alternatively, ones which will avoid some particular undesirable behaviour. The 
effects of any perturbation to the system's heterogeneity could then be pre-engineered in such a system. One possible future application would be to attack the global 
control problem of discrete actuating controllers\cite{AIAA}. We will also pursue our goal of tailoring multi-agent model systems to replicate the behaviour of a range of real-world systems, with a particular focus on (1) biological and human health systems such as cancer tumours and the immune system, and (2) financial markets.

\end{document}